\documentclass[twocolumn]{aastex63}

\usepackage{booktabs, multirow} % for borders and merged ranges
\usepackage{xcolor,colortbl}
\usepackage{changepage,threeparttable} % for wide tables
\usepackage{verbatim} 
\usepackage{gensymb}
\usepackage{svg}
\usepackage{shortbold}
\usepackage{xspace}
\usepackage{natbib}
\usepackage{amsmath}
\DeclareMathOperator*{\argmax}{arg\,max}
\DeclareMathOperator*{\argmin}{arg\,min}

\newcommand{\gauss}{Mx/cm$^{2}$\xspace}
\definecolor{darkgreen}{HTML}{008800}
\newcommand{\diff}[1]{#1}

\revised{December 15, 2020}
\accepted{January 2, 2021}
\submitjournal{ApJ}

\shorttitle{Fast Emulation of the SDO/HMI Pipeline}
\shortauthors{Higgins et al.}

\graphicspath{{./}}

\begin{document}

\title{Fast and Accurate Emulation of the SDO/HMI Stokes Inversion with Uncertainty Quantification}

\correspondingauthor{Richard Higgins}
\email{relh@umich.edu, fouhey@umich.edu}

\author[0000-0002-6227-0773]{Richard E. L. Higgins}
\affiliation{University of Michigan, Department of Electrical Engineering and Computer Science,
Ann Arbor, MI}

\author[0000-0001-5028-5161]{David F. Fouhey}
\affiliation{University of Michigan, Department of Electrical Engineering and Computer Science,
Ann Arbor, MI}

\author{Dichang Zhang}
\affiliation{University of Michigan, Department of Electrical Engineering and Computer Science,
Ann Arbor, MI}

\author[0000-0003-0176-4312]{Spiro K. Antiochos}
\affiliation{NASA GSFC,
Silver Spring, MD}

\author[0000-0003-3571-8728]{Graham Barnes}
\affiliation{NorthWest Research Associates Boulder,
Boulder, CO}

\author[0000-0001-9130-7312]{Todd Hoeksema}
\affiliation{Stanford University,
Stanford, CA}

\author[0000-0003-0026-931X]{K D Leka}
\affiliation{NorthWest Research Associates Boulder,
Boulder, CO}

\author[0000-0002-0671-689X]{Yang Liu}
\affiliation{Stanford University,
Stanford, CA}

\author[0000-0003-1522-4632]{Peter W Schuck}
\affiliation{NASA GSFC,
Silver Spring, MD}

\author[0000-0001-9360-4951]{Tamas I Gombosi}
\affiliation{University of Michigan,
Department of Climate and Space,
Center for Space Environment Modelling,
Ann Arbor, MI}

\begin{abstract}

The Helioseismic and Magnetic Imager (HMI) onboard NASA's Solar Dynamics Observatory (SDO) produces estimates of the photospheric magnetic field which are a critical input to many space weather modelling and forecasting systems. The magnetogram products produced by HMI and its analysis pipeline are the result of a per-pixel optimization that estimates solar atmospheric parameters and minimizes disagreement between a synthesized and observed Stokes vector. In this paper, we introduce a deep learning-based approach that can emulate the existing HMI pipeline results two orders of magnitude faster than the current pipeline algorithms. Our system is a U-Net trained on input Stokes vectors and their accompanying optimization-based VFISV inversions. We demonstrate that our system, once trained, can produce high-fidelity estimates of the magnetic field and kinematic and thermodynamic parameters while also producing meaningful confidence intervals. We additionally show that despite penalizing only per-pixel loss terms, our system is able to faithfully reproduce known systematic oscillations in full-disk statistics produced by the pipeline. This emulation system could serve as an initialization for the full Stokes inversion or as an ultra-fast proxy inversion.  This work is part of the NASA Heliophysics DRIVE Science Center (SOLSTICE) at the University of Michigan, under grant NASA 80NSSC20K0600E, and has been open sourced. \footnote{Find the code here: \href{https://github.com/relh/FAE-HMI-SI}{https://github.com/relh/FAE-HMI-SI}}
\footnote{Find the project site here: \href{https://relh.github.io/FAE-HMI-SI/}{https://relh.github.io/FAE-HMI-SI}}

\end{abstract}


\section{Introduction}
The Sun's magnetic field is the energy source for all solar activity, including energetic events such as flares and coronal mass ejections, that drive the most extreme space weather phenomena. Accordingly, there are multiple instruments measuring the photospheric magnetic field at a variety of duty cycles, spatial resolutions, and temporal cadences. One such instrument is the {\it Helioseismic and Magnetic Imager} \citep[HMI;][]{Schou2012} on the {\it Solar Dynamics Observatory} \citep[SDO;][]{PesnellThompsonChamberlin2012}, the first space-based instrument that produces  full-disk maps of the photospheric vector magnetic field  with a cadence of order minutes and a spatial resolution of order one arcsecond. HMI has collected almost a full, 11-year solar activity cycle's worth of consistently-acquired data, covering both maximum and minimum. Consequently, the HMI data have become the ``go-to'' for science investigations of solar activity by researchers throughout the world.   These data products are also used throughout the space weather community in applications ranging from flare forecasting research \citep[e.g.,][]{Bobraetal2015,DAFFS} to setting boundary conditions for coronal mass ejection modeling \citep[]{van2014alfven}.

Observational instruments do not directly measure the magnetic field. Instead, the magnetic field is estimated by first observing polarized light in a magnetically-sensitive spectral line, then modelling the photospheric plasma conditions that would best produce spectra consistent with those observed. With this method, instruments record polarized light (using the Stokes formalism) at multiple wavelengths before inverting the generative process and mapping photospheric parameters to polarized light. For instance, \textit{SDO}/HMI first measures six polarization states and transforms these observations to the four Stokes components $I$, $Q$, $U$, and $V$ at six wavelength positions (24 measurements per pixel). A subsequent algorithm, the Very Fast Inversion of the Stokes Vector \citep[VFISV;][]{borrero2011vfisv}, then inverts these Stokes vectors to produce magnetic and atmospheric parameters. VFISV forward models 8 parameters describing the magnetic field, kinematic, and thermodynamic properties in the photosphere with a Milne-Eddington (ME) atmosphere \citep{Unno1956,Rachkovsky62} to synthesize an estimated Stokes vector. VFISV then inverts this generative process by iterating the photospheric parameters with a Levenberg-Marquardt algorithm \cite[]{levenberg1944method, marquardt1963algorithm} until the discrepancy between the synthesized and observed Stokes vectors is minimized. This process results in 8 observables and associated uncertainties, namely: the field strength $B$, the plane-of-sky inclination $\gamma$ and azimuth $\Psi$, the line-of-sight component of the velocity of the magnetized plasma, the Doppler width, the line-to-continuum ratio $\eta_0$, the source continuum $S_0$, and the source gradient $S_1$.  \diff{Due to data limitations, the HMI pipeline VFISV does not invoke the magnetic fill-fraction parameter in the optimization, and assigns it to unity throughout \citep{Centeno2014} such that the returned field strength parameter $B$ is physically the pixel-averaged magnetic flux density \cite[]{graham2002inference}.}

The VFISV processing, like the majority of inversions of solar polarization data, are performed using a pixel-independent approach. Despite extensive optimization of the code and system-specific simplifications \cite[]{Centeno2014}, the analysis takes approximately 30 minutes using 8 cores per full-disk measurement computed with $4096 \times 4096$ pixels. Similar systems, such as MERLIN \cite[]{Lites2006} used in the Hinode/Solar Optical Telescope-SpectroPolarimeter data pipeline \cite[]{hinode, hinode_sp} are similarly slow.

Our paper presents a deep-network-based approach that provides an ultra-fast \diff{(4.8s per target} on a consumer GPU) emulation of this Stokes-inversion pipeline. This emulation encompasses both the Milne-Eddington model of the atmosphere and, importantly, the details of the HMI instrument specifications, i.e., instrument calibration, transmission profiles, measurement noise, etc.,
\diff{making it comparable to work like} \cite{ramos2019stokes}, which learns to invert from magneto-hydrodynamic simulations \diff{that were degraded using instrument spectral profiles and a spatial point spread function}.
This approach takes images containing, per-pixel, all the Stokes vector inputs (as well as metadata) and maps them to per-pixel estimates of a ME parameter produced by VFISV. \diff{Prior work has brought Stokes-vector inversion methods to the GPU \cite[]{harker2012gpu} with the goal of achieving real-time inversions, which our method achieves. We see our deep-learning approach as complementary to the porting of optimization-based methods to GPUs, and believe it has a number of benefits. First, the deep learning system requires only {\it samples} of inputs and outputs, which may reduce  effort needed to work on a new inversion scenario. Second, the framing as a deep network enables further acceleration via the extensive efforts towards task-independent acceleration of deep network forward passes, which range from reduced-precision arithmetic to quantization \cite[]{Jacob2018quantization}.} 

We base our approach on a U-Net \cite[]{ronneberger2015unet} architecture with a few crucial problem-specific modifications. This general approach has been used in other solar physics works such as by \cite{Galvez2019,Park2019} for \textit{SDO}/AIA UV/EUV image generation from \textit{SDO}/HMI magnetograms, and is a standard formulation used in areas such as biomedical image segmentation \cite[]{ronneberger2015unet}, pixel labeling \cite[]{shelhamer2017fully}, and general image translation \cite[]{pix2pix2017}.
Our proposed network is trained to solve the problem via regression-by-classification, where we train the network to match a distribution over a set of bins, which has been successfully used in computer vision for 3D prediction \cite[]{Ladicky14b,Wang15} and human pose estimation \cite[]{guler2018}, in part due to how it represents uncertainty compared to a more standard regression formulation. We demonstrate that this approach is capable of producing scientifically useful confidence intervals for all predictions.

The approach is trained on pairs of inputs and outputs from the existing pipeline's VFISV and thus aims to accurately emulate the results of the \textit{SDO}/HMI inversion rather than improve them.  Our approach is therefore most similar to recent work \cite[]{liu2020inferring} that emulates Stokes inversions on data from the Near InfraRed Imaging Spectropolarimeter (NIRIS) on the 1.6m \textit{Goode Solar Telescope} (\textit{GST}) at the Big Bear Solar Observatory, as well as work on fast learned inversion of differential emission measurements from \textit{SDO}/AIA data by \cite{Wright2018}. Our work differs across a number of crucial dimensions: our input measurements have far more limited spectral sampling (6 for \textit{SDO}/HMI vs 60 for \textit{GST}/NIRIS); methodologically, our proposed system can generate useful confidence intervals that communicate uncertainty; finally, we evaluate performance not only in terms of average per-pixel accuracy but also in trends over time in comparison to known system behavior.

% --- --- Pipeline --- ---
\begin{figure*}[t!]
\includegraphics[width=\linewidth]{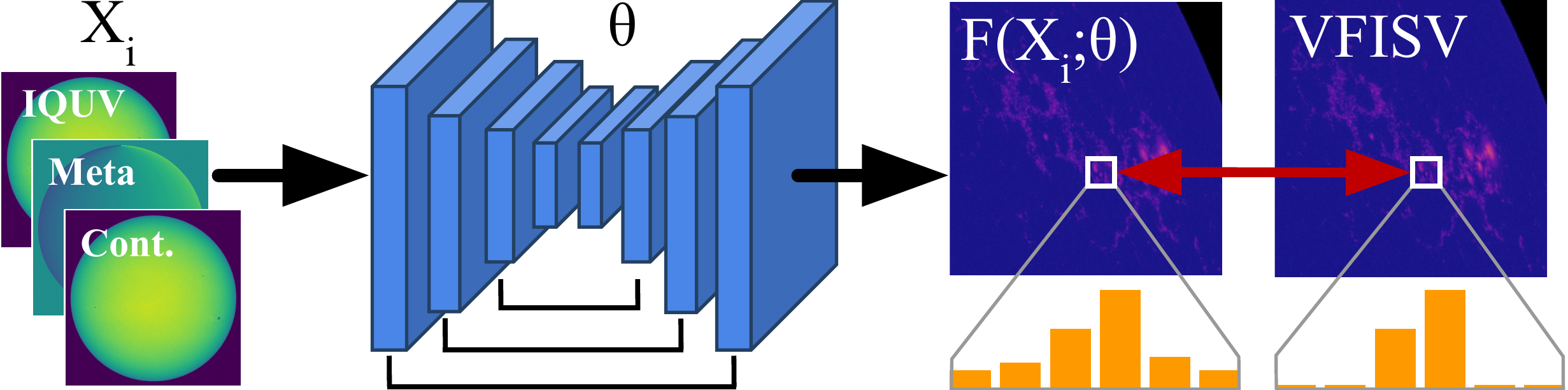}
\caption{Our approach for emulating the \textit{SDO}/HMI Stokes vector inversion pipeline, VFISV. As input, our network takes Stokes vector measurements (IQUV), metadata, and estimates of the continuum intensity (Section \ref{sec:inputarch}). As output it produces a per-pixel estimate of a single parameter of \diff{the inversion as would be produced by VFISV, e.g., inclination} (Section \ref{inference}). We cast the problem as regression-by-classification over a discrete set of bins. We show that this is both accurate (Section \ref{standalone accuracy analysis}) and enables fast uncertainty quantification (Section \ref{standalone confidence analysis}).}
\label{fig:pipeline}
\end{figure*}
% --- --- Pipeline --- ---

We evaluate how well this system can faithfully emulate VFISV in the \textit{SDO}/HMI pipeline via a series of experiments. \diff{We first train instances of the system, i.e., fit parameters of the neural network, on solar disks sampled from the first 60\% of 2015. We then validate the model's performance on data never encountered during training, sampled from the remaining 40\% of 2015. Finally, all evaluation, metrics, figures, and results are calculated and produced from test data consisting of solar disks sampled from the entirety of 2016. This test data is previously unseen to the model and separated in time from the training data by over 4 months. By employing temporally separate data regimes for training and testing, we ensure a fair evaluation of the proposed system.}

 \label{sec:intro}
\section{Methods}
We train a convolutional neural network \diff{(CNN) }to map observations of polarized light and auxiliary signals to estimates of a particular parameter of the photospheric magnetic field. Our network is a U-Net \cite[]{ronneberger2015unet}, an architecture capable of producing high-resolution per-pixel outputs that still consider and factor in a supporting spatial extent. As output, we modify this network to produce a distribution over a set of discrete values a magnetic field parameter can take; this distribution can be decoded via expectation into a single continuous estimate and additionally produce confidence bounds. To avoid tuning over eight simultaneous losses by handling all the parameters estimated by the SDO/HMI pipeline's VFISV inversion in a single network, we instead train individual networks for each of eight parameters used for ME Stokes vector generation. Training simultaneously over 8 targets could be done, but is not required given the considerable size of the dataset. We do not explicitly use the uncertainties produced by the pipeline's VFISV inversion.

\subsection{Input and Architecture}
\label{sec:inputarch} 

As input, our network takes a 28-channel image consisting of: 24 channels of the four components of the Stokes vector ($\IB, \QB, \UB, \VB$) observed in 6 passbands (the {\tt hmi.S\textunderscore720s} series); 1 channel of the continuum image (the {\tt hmi.Ic\textunderscore720s} series); and 3 auxiliary channels, comprising the heliographic coordinates computed via SunPy and a channel that is 1 if the pixel is off-disk and and 0 otherwise. \diff{No further calibration is performed on these published datasets.}

This input is passed through a U-Net-style convolutional neural network (Figure \ref{fig:pipeline}) that maps this 28-channel image to an 80-channel same-sized image, where each pixel encodes a distribution over possible outputs. 

This network is the composition of two jointly-trained subnetworks: an encoder that maps the input image to a smaller resolution image followed by a decoder that progressively maps this smaller resolution image back to the original size, mirroring the downsampling in the encoder with equivalent upsampling, finally resulting in an equal resolution output. As the decoder progresses, it incorporates information from both an upsampled version of the previous layer of the decoder as well as the equivalently sized image from the encoder via a skip-connection. In this instance, a skip-connection is an alternate path for information flow through the neural network, with connections between the encoder and decoder at layers of equal resolution. These skip-connections are represented by the black connectors in Figure \ref{fig:pipeline}. Combined, this architecture enables the decoder's inference to depend both on pixel-accurate local information (via the skip-connection) and broader context (via the encoder). The local information is important for precise per-pixel estimates and the broad context aids the inversion by enabling easy recognition of structures via shape. Though the VFISV pipeline does not consider spatial context across neighboring spectra, we include spatial context in our method to enable adjacency consideration as a potential information source when given a confusing input.

We largely follow standard design. The encoder and decoder blocks function similarly: after either a $2\times 2$ max-pooling for downsampling or a $2\times 2$ transpose-convolution for upsampling, there are two $3\times 3$ stride-1 convolutions (with each followed by a rectified linear unit (ReLU) \cite[]{nair2010rectified}. Spatial max-pooling is the process of only forwarding the largest output in a given height by width window to deeper layers for a downsampled output, while transposed convolution is a method of using learned weights to project a lower resolution image to higher resolution. Each halving/doubling of spatial resolution is accompanied by a doubling/halving of feature channels, and each convolution is zero-padded to preserve spatial resolution. The encoder has 4 downsampling blocks, which are mirrored by corresponding upsampling blocks. The first and last convolutional layers of the U-Net are not symmetric, respectively mapping the 28 input channels and 80 output channels to and from a 64 channel representation.  

One crucial difference is the omission of batch-norm. Batch normalization is a neural network component that calculates a mean and variance for inputs at a certain depth in the network and normalizes these inputs to help further computation later in the network. In addition to a nonlinearity like a ReLU, many CNNs conventionally interleave batch-normalization layers \cite[]{ioffe2015batch} that aim to statistically whiten (mean 0 and unit variance) each dimension of the internal features within a batch. We found it degraded model performance; we discuss likely causes of this in Section \ref{sec:experiments_ablations}.

In total, our network has 15 million trainable parameters. Our code and weights are publicly available at \href{https://github.com/relh/FAE-HMI-SI}{https://github.com/relh/FAE-HMI-SI}. Our code is implemented in PyTorch \cite[]{paszke2019pytorch}.

\subsection{Inference} \label{inference}
We cast our problem as a prediction of a distribution over a set of discrete values $\vB \in \mathbb{R}^K$ (throughout we set $K=80$ due to memory considerations). In particular, the network predicts at each pixel, a K-dimensional vector (of logits) that is converted to a distribution via the softmax function $\sigma(\zB_i) = \exp(\zB_i) / \sum_{j} \exp(\zB_j)$ \cite[]{bridle1990probabilistic}. After the softmax function, each pixel of output is a distribution $\hat{\yB} \in [0,1]^{K}$ or $\sum_j \hat{\yB}_j = 1$. This makes $\hat{\yB}$ an 80 element vector of probabilities. The bin values are linearly spaced depending on a range per inversion-output: for instance, for angles that range from $[0,180]$, we set $\vB_j = 180\times j/(K-1)$. 

To get a single value from this distribution, we decode a continuous value from the distribution. The most likely bin value is $\vB_m$ where $m = \argmax_j \hat{\yB}_i$. Following \cite{Ladicky14b}, we take an expectation over the adjacent values of this most-likely bin, or: 
\begin{equation}
\left( \sum_{j=m-1}^{m+1} \vB_j \hat{\yB}_j \right) \bigg/ \left( \sum_{j=m-1}^{m+1} \hat{\yB}_j \right).
\end{equation}
This scalar output is the final per-pixel prediction of our network. \diff{Though this approach produces continuous predictions, the propensity for convolutional neural networks to over-confidently predict a single bin \cite[]{pereyra2017regularizing, guo2017calibration} can lead this method of expectation to still have discrete, ``striped'' patterns in outputs, as shown in Figure~\ref{fig:bivariatehistograms}.}

One can similarly obtain a confidence interval $(l_l,l_u)$ by identifying the bin value at which the cumulative sum first exceeds a fixed threshold $\alpha$, or the $l$ where $\alpha = \sum_{j=1}^l \hat{\yB}_{j}$. This can be calculated with sub-bin accuracy by linearly interpolating the cumulative distribution function (CDF) between bins. Thus, assuming the output is at the median, one obtains a 90\% confidence interval (CI) by solving for $l_l$ and $l_u$ for which the CDF is $5\%$ and $95\%$ respectively.

In practice, neural networks tend to have poorly calibrated confidence intervals and we therefore re-calibrate the interval on held-out data (not used in evaluation) by fitting two simple correction factors. The first, following \cite{neumann18relaxed}, incorporates a temperature $\tau$ in the softmax, or 
$\sigma(\zB,\tau)_i = \exp(\tau \cdot \zB_i) / \sum_j \exp(\tau \cdot \zB_j)$
where $\tau \to 0$ softens the distribution to a uniform distribution and $\tau \to \infty$ sharpens it to a one-hot (1 in the correct class location and 0 elsewhere) vector. One can fit $\tau$ to ensure the empirical confidence interval covers $90\%$ of the data. This improves results, but for relatively wide intervals, we find that the intervals can still underestimate since they do not cover enough of the output space to sufficiently account for outliers. We additionally apply a simple and empirically effective post-hoc correlation where we expand the interval $(l_l,l_u)$ around the center by a factor of $\beta$. We compute this $\beta$ as the ratio between the target coverage (e.g., 90\%) and the empirical coverage on held-out data. 

\subsection{Objective and Training}

The network is trained on $N$ pairs of inputs and corresponding targets $\{\XB^{(i)},\YB^{(i)}\}$, with height $H$ and width $W$. Suppose
$f: \mathbb{R}^{H \times W \times 28} \to \mathbb{R}^{H \times W \times 80}$ is the function that the convolutional neural network represents and $\thetaB$ represents all of the trainable parameters of the network (i.e., all of the convolution filter weights and biases for the encoders and decoders mentioned in Section \ref{sec:inputarch}).  We seek to solve the minimization problem:
\begin{equation}
\label{eqn:objective}
\argmin_{\thetaB}
\sum_{i=1}^N \sum_{p} 
 \mathcal{L}(f(\XB^{(i)};\thetaB)_{p},\YB^{(i)}_{p}),
\end{equation}
or the minimization with respect to $\thetaB$ of the sum, over each pixel $p$ in each image $i$, of a loss function measuring how well the estimate $f(\XB^{(i)};\thetaB)$ matches the target $\YB^{(i)}$.

Given an estimate of the distribution $\hat{\yB}=f(\XB^{(i)};\thetaB)_p$ and target $y = \YB^{(i)}_p$, we penalize estimates $\hat{\yB}$ that deviate from a target discrete distribution $\dB$ that has $y$ as its expected value. In particular, $\dB$ is created by identifying the subsequent bin values $v_b$ and $v_{b+1}$ that bracket $y$ (above and below) and then solving for probabilities $\dB_b$ and $\dB_{b+1}$ that make $y = \dB_b v_b + \dB_{b+1} v_{b+1}$. To also discourage network overconfidence, other bin values take on a small probability of $10^{-4}$. The final error is the Kullback Leibler (KL) divergence \cite[]{kullback1951information} which measures the deviation  between $\hat{\yB}$ and $\dB$, or 
\begin{equation}
\mathcal{L}(\hat{\yB},\dB) = - \sum_{j=1}^K \dB_j \log\left(\frac{\hat{\yB}_j}{\dB_j} \right).
\end{equation}
Overall, this approach penalizes an inaccurate final estimate by penalizing the difference between a predicted probability distribution across bins and a constructed one.

An alternative that we explored, and compare quantitatively to, is minimizing the divergence from $\hat{\yB}$ to a one-hot distribution where only the nearest bin $m$ had probability mass. This is equivalent to the standard negative log-likelihood loss, or $-\log(\yB_m)$. In practice, we found that smoothed encoding produced superior results. We hypothesize that this is because the divergence to one-hot encoding encourages predicting the nearest bin rather than producing a distribution: in low-field strength regions, for instance, if the network can identify that the field is nearer to zero than the first bin-value, then it is rewarded for placing its probability mass in the zeroth bin rather than producing a distribution. Since many of the targets do not follow a uniform distribution of values, we experimented with weighting the probabilities in the KL divergence loss with the inverse frequency of each bin (along with a bias to prevent enormous weights when taking the inverse). Although weighting helped training in the negative log likelihood setting, it did not improve performance.

We note that irrespective of the loss function at each pixel, the total training objective is per-pixel in the sense that it is a sum of one term per pixel with no terms that tie together pixels (e.g., ensuring that summary statistics between the network and the ground truth match). Thus, while the network is not per-pixel on the input size since each output pixel depends on a set of pixels on the input side, the network has no indication that its goal should include anything beyond matching each pixel independently and as closely as possible. This leaves room for various improvements, although our experiments show that the network trained as it is reproduces some important summary statistics.

We solve for the network parameters of each model by minimizing Equation \ref{eqn:objective} with respect to $\thetaB$ via stochastic gradient descent \cite[]{robbins1951stochastic}, using the AdamW optimizer \cite[]{loshchilov2017decoupled}, with learning rate $10^{-4}$, $\epsilon=10^{-4}$, and weight decay $3\times10^{-7}$. Optimization scheduling was accomplished by monitoring loss on held-out data: the learning rate was halved if there were two consecutive epochs without validation loss and terminated if there were four. Validation data were used to fit $\tau$ and $\beta$ for CI calibration with half used to fit $\tau$ and half to fit $\beta$.

\subsection{Speed and Implementation Details}

Due to GPU memory constraints, each $4096\times4096$ full-disk image is divided into $16$ $1024\times1024$ pixel tiles. On \diff{a GeForce RTX 2080 Ti GPU with 4352 CUDA cores, }we find that inference on each tile takes on average 300 ms once data has been loaded into main memory. In running the full system, the time spent loading from disk is the primary bottleneck, even when using solid-state drives. Running all 16 tiles sequentially on a single GPU thereby takes 4.8 seconds. This time includes the time spent to load input from main memory to GPU memory, the time spent running the neural network on this input, and finally the time taken to turn output probabilities into regressed values. \label{sec:methods}
% --- --- Qualitative Results --- ---
\begin{figure*}
\includegraphics[width=\linewidth]{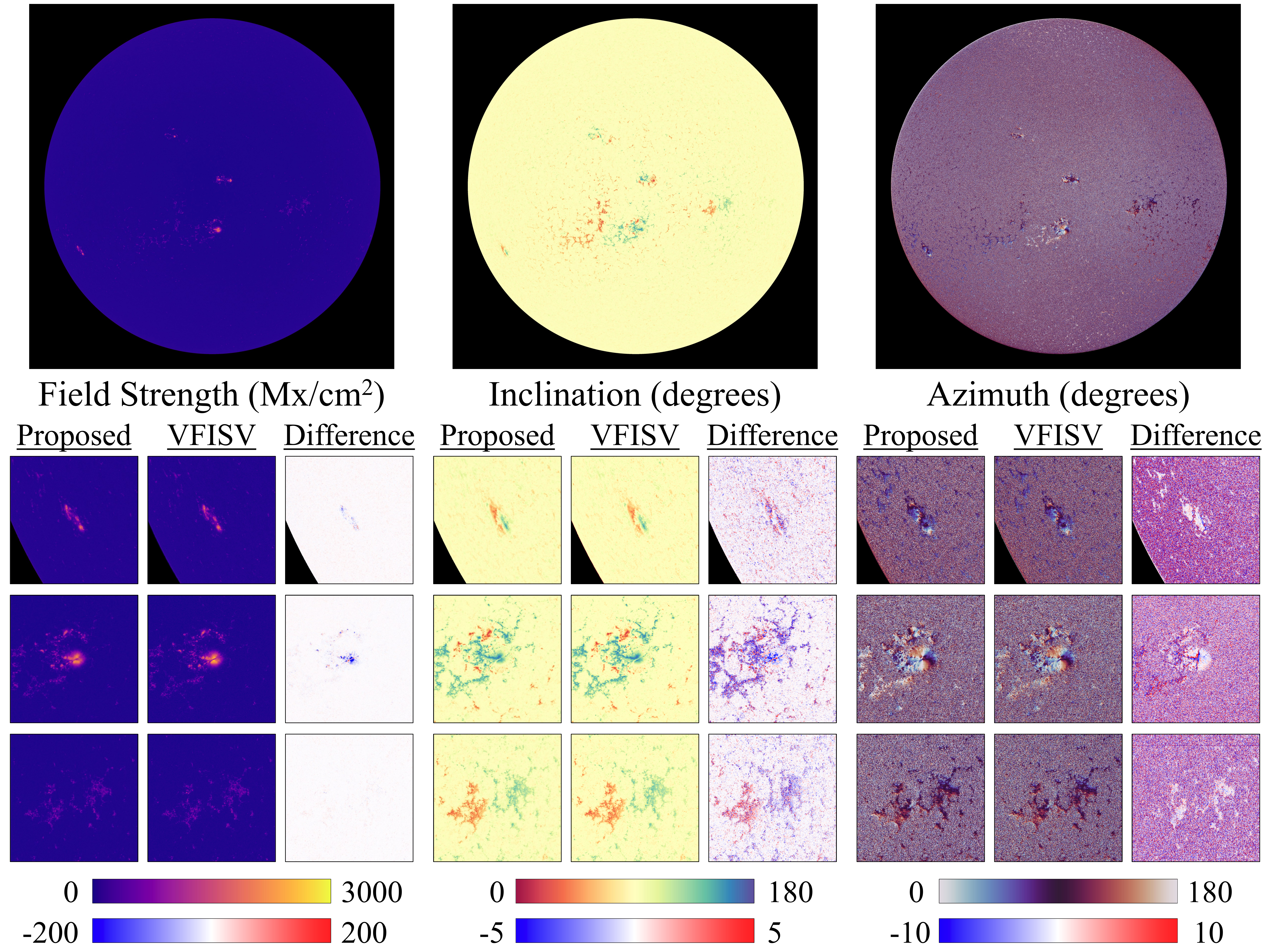}
\caption{Qualitative results for full disk and a few active regions on held-back data corresponding to observations at 2016/05/10-06:48:00\,TAI. The predicted full disk images for magnetic field strength, inclination, and azimuth are generated by the proposed approach. We show cutouts for a few different areas -- an active region towards the east limb, an active region in the center of the disk, and plage to the west of disk center. Field strength and inclination estimates are generally precise in regions of both moderate and weak polarization. Azimuth, on the other hand, is poorly constrained in areas of weak linear polarization. However, the azimuth-angle predictions from VFISV are also poorly constrained in such areas, therefore it is consistent that the proposed emulation method is similarly noisy.}
\label{fig:fulldisk}
\end{figure*}
% --- --- End Qualitative Results --- ---

\section{Experiments}

We conduct a series of experiments to quantitatively answer a number of questions about the proposed emulation technique. From the start, we stress that our goal is to {\it emulate} the \textit{SDO}/HMI and VFISV pipeline with high fidelity, rather than to improve it. In particular we assess: (a) how accurately we emulate the current pipeline on {\it held-out} data, what parts of the data are particularly well-emulated (and which are not), and whether our confidences correlate with uncertainties produced by the existing pipeline; (b) how well the proposed approach compares to alternative approaches, including using more standard regression, the network originally used to initialize VFISV, and a network using batch-norm; (c) how the performance varies in the temporal domain in addition to static single-snapshot evaluations, namely whether we emulate (for example) the known 24-hour periodic oscillations in the pipeline.

\begin{figure*}[t]
\newcommand{\qualwidth}{0.18\linewidth}
\centering
\includegraphics[width=\linewidth]{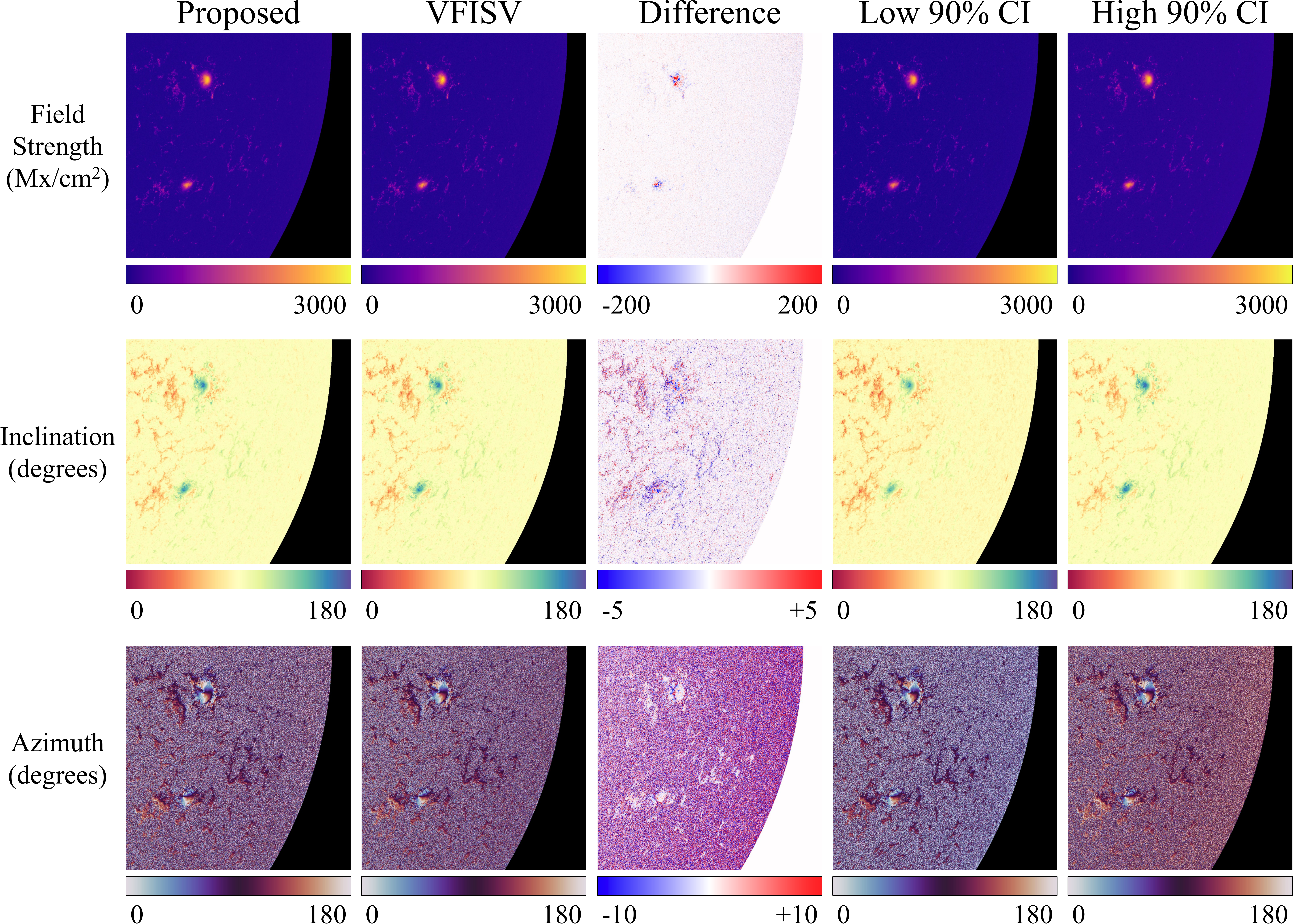}
\caption{Performance compared to ground truth results for field strength \diff{(magnetic flux density)}, inclination, and azimuth on held-back, unseen data from 2016/01/15 at 12:36:00\,TAI. The lower and upper bounds of the 90\% confidence interval are also visualized. The similar lower and upper bounds for \diff{flux density} and most of \diff{the} pixels for inclination indicate the network is confident the true value is believed to lie in a narrow range. In \diff{the} active region inclination maps, one can see variations in color saturation but not hue across the lower and upper bounds, indicating confusion about the distance to $90^\circ$ but not direction.}
\label{fig:qual_main}
\end{figure*}

\begin{figure*}[t]
\includegraphics[width=\linewidth]{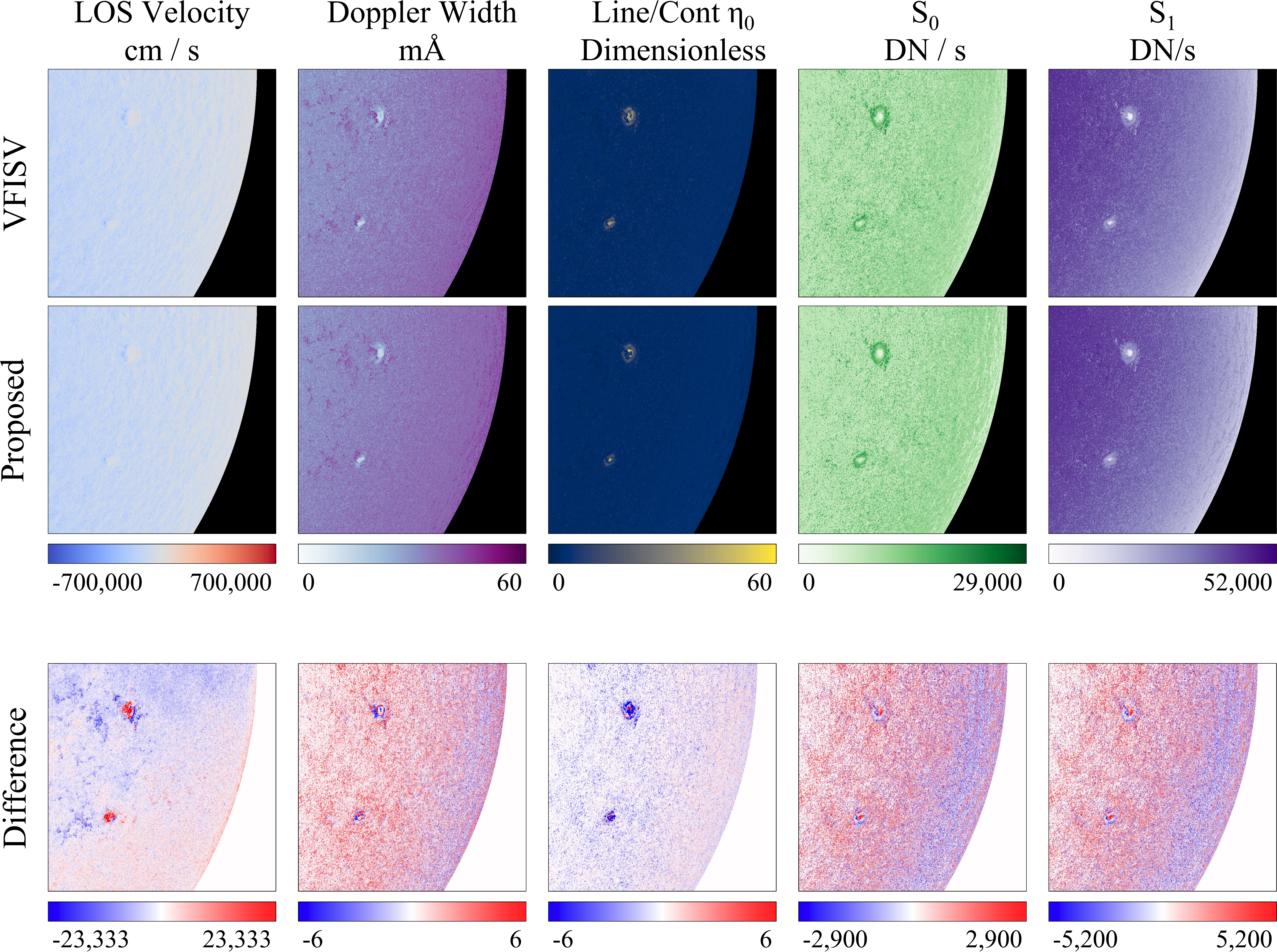}
\caption{Prediction of the Kinematic (LOS Velocity) and Thermodynamic parameters (Doppler Width, Line-to-Continuum Ratio $\eta_0$, Source Function Constant Term $S_0$, and Source Function Gradient Term $S_1$) on held-back, unseen data from 2016/01/15 at 12:36:00\,TAI.}
\label{fig:qual_thermo}
\end{figure*}

\begin{figure*}
\centering
\includegraphics[width=\linewidth]{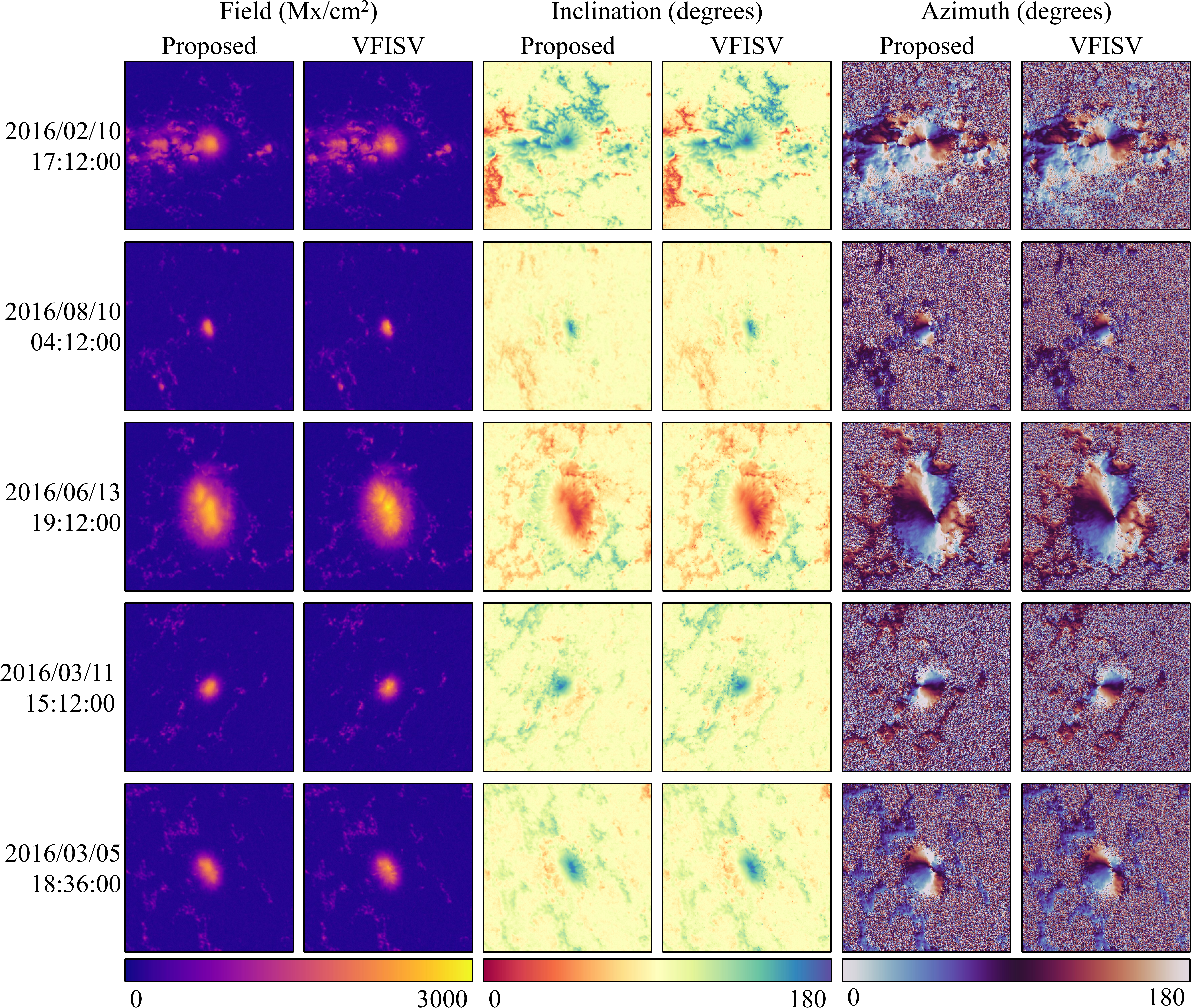}
\caption{Results on \diff{high-signal regions. We cutout $125^{\prime\prime}$-square regions} by {\it randomly} selecting test set dates from {\bf 2016-All}, cropping around the pixel with the highest field strength, and skipping dates if they appear elsewhere in the manuscript, the pixel was on-limb, or less than 1\% of the cutout had field strength more than 500 \gauss. There is some smoothing, which reduces speckling in inclination and some small loss of detail in field.}
\label{fig:random}
\end{figure*}

\subsection{Datasets}\label{sec:datasets}

Three sets of data from JSOC of the SDO/HMI data were employed, and will be given DOIs and made publicly available. In particular:
\begin{itemize}
    \item {\bf 2015-All:} A regular cadence every two days starting January 1, 2015 - 31 December 2015, sampling an hour and minute randomly (from cadences available) and ignoring any failures. We chronologically split this into a training, validation, and test set in 60/20/20\% proportion. The last observation in the training set and the first observation in the test set are separated by over two months. The training portion of this dataset is used for training and small-scale validation of our model.
    \item {\bf 2016-All:} We repeat the above procedure, but for 2016. These data are never used for training.
    \item {\bf 2016-Month:} To investigate whether the proposed approach successfully replicates known oscillatory behavior, we pick a random month from 2016 and sample hourly at 36 minutes past-the-hour. These data are never used for training.
\end{itemize}

\subsection{Outputs and Data Preparation}

We predict the eight magnetic and thermodynamic parameters produced by the pipeline's VFISV inversion and stored in the {\tt hmi.ME\textunderscore720s\textunderscore fd10} series. For each output, we identify a target range that our classification network predicts, which is determined by starting at the 99\% range of the data, adjusting for physical plausibility and important rare values; we also report the units as originally reported by VFISV, along with any important findings. In particular, we note quantities that are known to have unphysical 24-hour periodic oscillations. A more complete description can be found in \cite{Hoeksema2014}.

\begin{itemize}
    \item {\bf Field Strength} ($B$), \diff{albeit physically the magnetic flux density,} which we predict from $0$ to $5000$ \gauss.  \diff{Strong} values are rare but important to predict correctly. The average on-disk \diff{flux density} (i.e., $\frac{1}{n}\sum_{p} B_p$ where $p$ indexes over the $n$ on-disk pixels), dominated by inferred low-strength values, is known to oscillate with the orbital velocity (plus harmonics). 
    \item {\bf Inclination} ($\gamma$), which we predict from $0^\circ$ to $180^\circ$. There is a preferred direction of $90^\circ$, which dominates in low polarization regimes but is understood to be an influence of noise \diff{ \cite[]{borrero2011inferring}; }as polarization signals (both linear and circular) increase, accuracy increases however the precision of the prediction becomes worse, likely because because the inclination becomes more varied and high-polarization points are relatively rare. The average distance from $90^\circ$ (i.e., $\frac{1}{n} \sum_p |\gammaB_p - 90|$) also oscillates with SDO's orbit\diff{ see Figure \ref{fig:cat_ears} and \cite[]{Hoeksema2014}.}
    \item {\bf Azimuth} ($\Psi$), which we also predict from $0^\circ$ to $180^\circ$ (the $180^\circ$ ambiguous azimuth as per the output of the pipeline VFISV inversion).  In areas of low linear polarization, this value is less constrained and far noisier, making it more difficult for a network to predict.
    \item {\bf Line-of-Sight (Doppler) Velocity} ($v$), which we predict from $-700,000$ to $700,000$ $\textrm{cm}/\textrm{s}$. This velocity is with respect to the instrument, and thus accounts for the solar plasma velocity itself, solar rotation, and the orbital velocity of the instrument's satellite (imparted by maintaining a geosynchronous orbit for transmission to a ground station).
    \item {\bf Doppler Width} ($\Delta\lambda_{\rm D}$), which we predict from $0$ to $60$ m\AA.  This parameter is not well constrained in the pipeline VFISV due to degeneracy with other thermodynamic variables and, at some level, the Zeeman splitting itself.  The pipeline VFISV instituted a variable substitution to address this by simultaneously fitting $\Delta\lambda_{\rm D}$ and $\eta_0$ \citep{Centeno2014}, but the proposed method addresses each parameter separately.
    \item {\bf Line-to-Continuum Ratio} ($\eta_0$), which we predict from $0$ to $60$. This dimensionless quantity is not well constrained by the \textit{SDO}/HMI observations due to low spectral resolution and degeneracy with other thermodynamic variables \citep{Centeno2014}. The VFISV pipeline therefore has a regularization term that encourages solutions close to a constant, $5$.
    \item {\bf Source Function Constant} ($S_0$), which we predict from $0$ to $29,000$ data number (i.e., counts from the CCD) per second, or $\textrm{DN}/\textrm{s}$.
    \item {\bf Source Function Gradient} ($S_1$), which we predict from $0$ to $52,000$ $\textrm{DN}/\textrm{s}$. 
\end{itemize}
The magnetic field (strength and angles) and kinematic property (LOS velocity) additionally include uncertainties, which are computed as proportional to the inverse of the diagonal elements on the Hessian of the VFISV minimization objective, multiplied by the final $\chi^2$ objective function value\diff{, see Eq. 11.29 in \cite[]{del2003introduction}.}

The inputs to our model come from the {\tt hmi.S\textunderscore720s} and {\tt hmi.Ic\textunderscore720s} series as well as from calculations done by SunPy \cite{sunpy_community2020} on this data to obtain solar latitude/longitude. Throughout, we operate on images that have been rotated according to the {\tt CROTA\textunderscore2} FITS header parameter via SunPy. For those interested, we also offer models trained on the unrotated images.

\section{Results}
\subsection{Qualitative Results}

% --- --- Main Table --- ---
\begin{deluxetable*}{lc@{~~~~~~~~}ccccccccc}
\caption{Quantitative evaluation of VFISV emulation results across 8 Stokes vector inversion targets across the 2016-All dataset. We evaluate according to the Mean Absolute Error (MAE) and percent of pixels within t. Values for $t$ are target specific and are generated by scaling according to the relative variances. We report numbers on four populations of pixels, defined in-text: ({\bf Disk}) On-Disk, ({\bf Plage}) Plage Pixels, ({\bf AR}) Active Region Pixels, ({\bf 100+}) Pixels with at least 100 \gauss in the absolute value line of sight \diff{magnetic flux density.} } 
\label{tab:primary}
\tablewidth{0pt}
\scriptsize
\tablehead{\colhead{Target}&\colhead{Range}&\multicolumn4c{MAE}&\colhead{$t$}&\multicolumn4c{\% Within $t$}\\
&&\colhead{Disk} &\colhead{Plage}&\colhead{AR}&\colhead{100+}&&\colhead{Disk}&\colhead{Plage}&\colhead{AR}&\colhead{100+}}
%\decimalcolnumbers
\startdata
Field (B) &[0,$5\!\times\!10^3$] \gauss &9.67 &18.87 &108.44 &16.98 &47 &99.2\% &93.1\% &31.9\% &93.6\% \\
Inclination ($\gamma$) &[0,180] \degree &0.58 &2.42 &2.53 &2.46 &5 &98.9\% &88.1\% &84.1\% &87.1\% \\
Azimuth ($\Psi$) &[0,180] \degree &13.06 &9.58 &10.67 &8.58 &7 &59.9\% &75.1\% &71.2\% &77.9\% \\
LOS Velocity &[$-7\!\times\!10^5$,$7\!\times\!10^5$] cm/s &5,247 &7,797 &23,834 &7,010 &38,800 &99.7\% &99.4\% &80.6\% &99.0\% \\
Dop.~Width &[0,60] m\AA &1.36 &1.83 &6.29 &1.67 &0.96 &63.2\% &45.9\% &15.0\% &56.5\% \\
Line/C. $\eta_0$ &[0,60] &0.78 &0.77 &10.19 &1.00 &0.81 &79.4\% &81.1\% &11.0\% &82.0\% \\
SrcCont. $S_0$ &[0,$2.9\!\times\!10^4$] DN/s &969 &2,371 &2,494 &2,041 &813 &74.0\% &53.4\% &23.4\% &63.2\% \\
SrcGrad. $S_1$ &[0,$5.2\!\times\!10^4$] DN/s &1,234 &2,592 &2,874 &2,218 &1,814 &80.7\% &66.7\% &41.2\% &73.4\% \\
\enddata
\end{deluxetable*}

We first show qualitative results on held-back data in Figure \ref{fig:fulldisk} with full disk results, as well as on the same tile in Figure \ref{fig:qual_main} for magnetic field parameters and Figure \ref{fig:qual_thermo} for kinematic and thermodynamic parameters. Overall, despite using discrete probabilities, banding patterns in the output are usually relatively difficult to identify.

\vspace{2mm}
\noindent {\bf Field Strength} \diff{parameter} is modeled qualitatively precisely (close to the VFISV output), overall. Results are difficult to distinguish by eye and require a relatively tight difference map ($\pm 200$ \gauss, or $4\%$ of the total range) to clearly bring out errors. The largest areas of frequent and significant error are in strong-field regions (see more discussion in Section \ref{sec:limitations}).

\vspace{2mm}
\noindent {\bf Inclination} is modeled well qualitatively, although some systematic errors exist. These errors can be identified in difference maps, where red corresponds to over-prediction and blue under-prediction: plage pixels with inclination $<90^\circ$ tend to be over-predictions in the difference map; pixels with inclination $>90^\circ$ tend to be under-predictions. 
This prediction uncertainty is reflected in the inclination confidence intervals (Fig.~\ref{fig:qual_main}) in the active regions: the difference in saturation between the lower and upper CI values, but not hue, shows that the network is sure of the direction (up-or-down) but less sure of how far up or down. 

\vspace{2mm}
\noindent {\bf Azimuth} is far noisier over much of the disk, where linear polarization ($[Q,~U]$) is low, so the difference map is difficult to interpret.  However, as these signals increase, the trained system does a qualitatively good job at modeling the complex patterns in the azimuth, as seen by good spatial correspondence and substantial regions of white (agreement) in the difference maps. Meanwhile, the VFISV-produced azimuths are highly random as well in noise-dominated areas, which explains why the difference map is so pronounced there  -- the network has understandable difficulty predicting random outcomes.

\vspace{2mm}
\noindent {\bf Line-of-Sight  (Doppler) Velocity} is generally estimated precisely, although less so in \diff{strong-polarization} regions. There is a noticeable change in the direction of error at the limb in quiet regions that occurs on many other dates as well. 

\vspace{2mm}
\noindent {\bf Thermodynamic parameters} are generally estimated with fair precision.
We note a trend of the network to oversmooth details in \diff{high-strength} regions (contrast, for instance, the small details in the \diff{sunspot} regions in VFISV compared to the proposed approach). Just as with the Doppler velocity, the source continuum and gradient change error modes closer to the limb and in quieter areas.

\begin{figure*}[t]
\centering
\includegraphics[width=\linewidth]{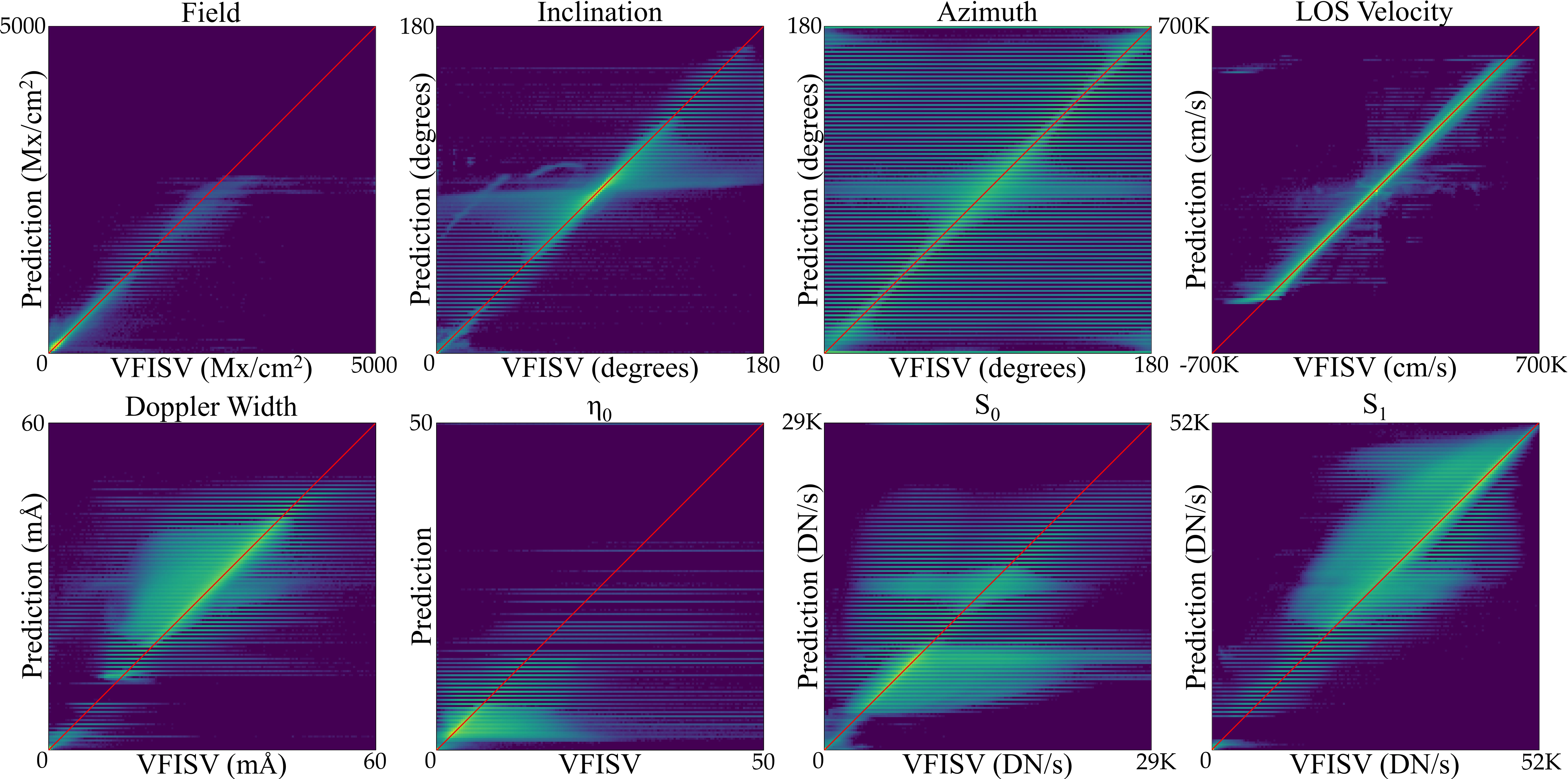}
\caption{Visualizing classification model prediction performance with {\bf log count} bi-variate histograms across the 2016-All data. Perfect prediction would put all data on the $y=x$ line (shown in red). While more rare bins or bad errors do have a banding effect (due to classification binning), most of the pixels lie along the $y=x$ line, with relatively good agreement with VFISV.}
\label{fig:bivariatehistograms}
\end{figure*}

\subsection{Standalone Accuracy Analysis} \label{standalone accuracy analysis}

Quantitative results are next reported for all of the outputs on the \textbf{2016-all} dataset (Table \ref{tab:primary}). These numbers are best interpreted alongside the per-output bivariate \diff{log} histograms in Figure~\ref{fig:bivariatehistograms}, where data off the $y=x$ line indicate the presence of estimation errors.

\vspace{2mm}
\noindent {\bf Metrics:} 
We quantify performance between a scalar target $y$ and inference result $\hat{y}$ with two styles of metrics. The first is the mean absolute error (MAE), or the average of $|y-\hat{y}|$ over the dataset. Since this is the average distance between the target and the inferred value, it can be highly influenced by outliers, \cite[]{Scharstein02}: a mean absolute  error of $20^\circ$ could be either all pixels being off by $20^\circ$ or 80\% of pixels being off by $4^\circ$ and 20\% being off by $84^\circ$. We therefore also compute percent-good-pixels or the average number of pixels satisfying $|y - \hat{y}|<t$ for a threshold $t$. To avoid picking eight independent thresholds, we pick a threshold for inclination angle ($5^\circ$ to represent reasonably close predictions) and then scale this for other variables. Mimicking how $R^2$, the coefficient of determination, scales by variance, we scale the thresholds by the relative variances (i.e., field strength 
$t_B / t_\gamma = \textrm{var}(B) / \textrm{var}(\gamma)$). We report the average for each quantity, thresholded at the appropriate $t$.

\vspace{2mm}
\noindent {\bf Pixel Populations:}
Evaluating error statistics over the full disk does not give a full picture of the results since the vast majority of on-disk pixels have low signal-to-noise. We therefore report per-pixel evaluations on the full disk as well as regions that aim to capture {\it Plage},  {\it Active Regions (AR)}, and pixels with at least 100 \gauss (100+). We define Plage pixels as any pixel with: continuum intensity (from {\tt hmi.Ic\_noLimbDark\_720s}) $\ge 0.8$; LOS absolute \diff{flux density} (from {\tt HMI.M\_720s} due to its reduced noise) $>100$ \gauss; and disambiguation confidence, {\tt conf\_disambig} (from {\tt hmi.B\_720s}) $\ge 60$. The Plage mask primarily includes Plage, but also includes a small amount of outer penumbra, and accounts for $\sim 0.4\%$ of the data in the 2016-All dataset. We define {\it Active Regions} using the same series, requiring continuum intensity $< 0.8$, LOS absolute \diff{flux density} $>100$ \gauss, and disambiguation confidence $\ge 60$. The Active Region mask accounts for $\sim 0.02\%$ of the data. Finally, we evaluate on pixels with at least $>100$ \gauss absolute value in the line of sight \diff{flux density}, which accounts for $\sim 47\%$ of the data.

\vspace{2mm}
\noindent {\bf Field Strength} \diff{parameter is difficult to model accurately because most pixels correspond to low intrinsic field strength or unresolved structures on the Sun, while pixels with both high intrinsic  field strength and large fill fraction (thus presumably resolved)} are relatively rare. Although our discretization steps are $63.3$ \gauss apart, the network is able to achieve sub-bin precision with an MAE of $9.67$ \gauss and with $99.2\%$ of the pixels within $47$ \gauss. In the rare regimes of $\gtrsim2750$ \gauss, the predicted output is generally underestimated, likely due to extreme data scarcity.

\vspace{2mm}
\noindent {\bf Inclination} is similarly well-estimated, with a MAE of $0.58^\circ$, around 25\% of the $2.27^\circ$ bin width. One might suspect that the low error in full disk prediction is driven by the positive-definite character of the noise in the transverse component of the field estimates \diff{leading to a preference for an inclination angle of $90^\circ$ especially in quiet-sun (low-polarization) regions.} Nonetheless, it is not required, as the approach still achieves a low MAE of $2.46^\circ$ in \diff{the 100+ regime,} with over $87.1\%$ of those values predicted within $5^\circ$. As seen in Figure~\ref{fig:bivariatehistograms}, the relatively few gross inclination errors are rarely on the wrong side of $90^\circ$ (seen by counts in the upper-left or bottom-right quadrant), but rather an underestimate of the angle magnitude.

\vspace{2mm}
\noindent {\bf Azimuth} has the opposite difficulty compared to inclination since it is noisy (and therefore difficult to estimate) in regions with weak linear polarization, which can occur in both strong- and weak-field regimes although the former is extremely rare, occurring in small areas within sunspots.  As overall polarization signal increases (into stronger-field regimes), the MAE improves, going down substantially from $13.06^\circ$ on-disk to $8.58^\circ$ in $>100$ \gauss LOS absolute \diff{flux density} pixels, and $\approx 10^\circ$ in the AR and Plage areas.

\vspace{2mm}
\noindent {\bf Line of Sight Velocity} has the tightest estimates of all the outputs in the bivariate histograms and nearly every pixel ($99.7\%$) falls within the threshold. This is in part because much of the variability is driven by a global parameter corresponding to the spacecraft's velocity at the time of data acquisition. As \diff{the field parameter (flux density)} increases, the error jumps substantially. However, as seen in Figure \ref{fig:bivariatehistograms}, sign-flips are relatively rare, as is the case for inclination.

\vspace{2mm}
\noindent {\bf Thermodynamic Parameters}, are predicted similarly well. Source Continuum, Source Gradient, Doppler Width, and $\eta_0$ are relatively precisely modeled and are roughly similar in behavior to inclination. $\eta_0$ prediction performance is not excellent, yet this behavior is consistent with the observations of \cite{Centeno2014} that there are degeneracies in the pipeline relations of field strength and $\eta_0$. The Doppler Width histogram shows an interesting separation of values, where almost no predictions land in the interval between 8 and 13 m\AA\diff{, as seen in the bottom-leftmost panel of Figure \ref{fig:bivariatehistograms}.} 

Across all targets prediction quality is good, with low average error. \diff{Despite our method producing continuous output values, we find that the bivariate log histogram in Figure \ref{fig:bivariatehistograms} reveals a banding pattern as a product of overly confident predictions in regression-via-classification. This is, however, usually less pronounced near the $y=x$ line.} Of all targets, azimuth angle is the only one to improve absolutely for high field strength \diff{(nominally high polarization) regions}. The absolute performance of other targets decreases in ARs, however, their relative/fractional error may actually be smaller than other pixel populations. 

\subsection{Standalone Confidence Analysis} \label{standalone confidence analysis}

We next analyze the performance of our network at predicting upper and lower confidence bounds for where pipeline VFISV outputs will fall, on the \textbf{2016-all} dataset. We do this by comparing our uncertainties with absolute error (i.e., is the approach less accurate on pixels that it is less certain about?) and with pipeline uncertainties (i.e., is the approach generally more uncertain about the same pixels as the generating pipeline VFISV output?). 

\begin{figure}[t]
\includegraphics[width=\linewidth]{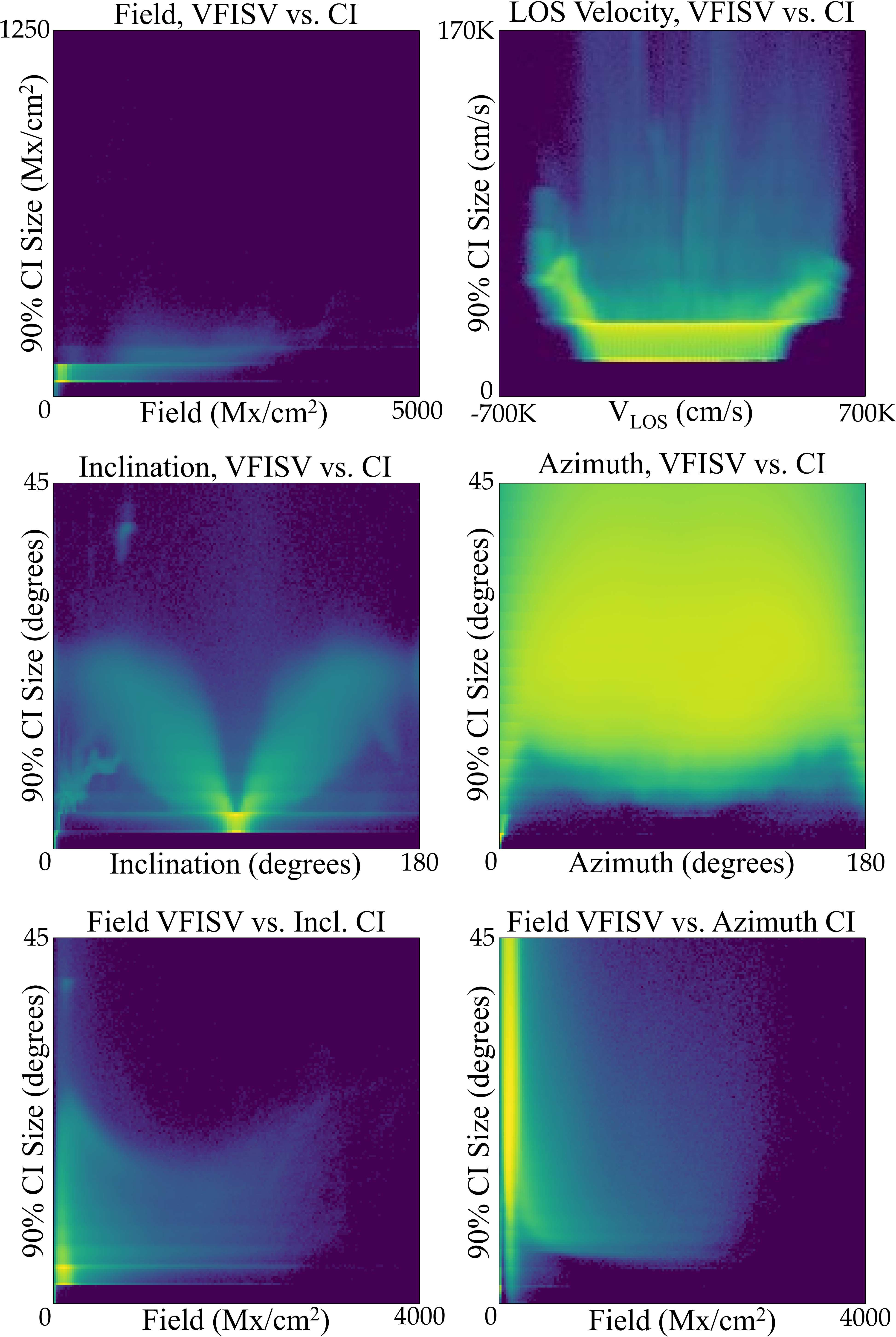}
\caption{Log histograms comparing the width of the network's 90\% confidence interval to ground truth values for field strength, LOS velocity, inclination, and azimuth (rows 1-2). Row 3 compares the width of the network's 90\% confidence interval against field strength for both inclination and azimuth angle. Results calculated on all pixels in the \textbf{2016-all} dataset.}
\label{fig:widthvalue}
\end{figure}

We begin with some qualitative behavior in Figure~\ref{fig:widthvalue} which shows how parameter values and confidence interval width co-vary.
Inclination prediction is less confident as it deviates from $90^\circ$, as shown by the two plumes, matching the qualitative behavior seen in identical lower and upper bounds in quiet sun. Quiet sun (noise-dominated) azimuth, on the other hand, is uniformly distributed, and therefore there is close to no relationship between azimuth and width. 
Finally, field strength and Doppler velocity are unsurprisingly more confident closer to zero and prediction widths increase as magnitude increases.

\begin{deluxetable}{ccccc}[t]
\caption{Spearman's $\rho$ between Emulated Confidences and absolute error, classification bin width, average width of 90\% confidence interval, and percent of test set within the 90\% confidence interval measured on 2016-all. While some $\rho$ are low, all are reported as significant. Many of the test set intervals are close to including $90\%$ of the data, although some are under- or over-estimated.}
\label{tab:conferror}
\tablehead{\colhead{Target} & \colhead{$\rho$} &
\colhead{Bin Width} & \colhead{CI Width}  & \colhead{\% in CI}}
\startdata
B (\gauss) &0.04 &63.3 &64.4 &88.5\% \\
$\gamma$ (\degree) &0.32 &2.3 &4.2 &86.1\% \\
$\Psi$ (\degree) &0.47 &2.3 &51.3 &90.2\% \\
$v_{\textrm{los}}$ (cm/s) &0.15 &17,721 &31,992 &77.5\% \\
$\Delta \lambda_D$ (m\AA) &0.42 &0.76 &2.2 &60.0\% \\
$\eta_0$ &0.42 &0.76 &0.97 &66.5\% \\
$S_0$ DN/s &0.63 &367 &1049 &60.3\% \\
$S_1$ DN/s &0.57 &658 &5565 &88.5\% \\
\enddata
\end{deluxetable}

To answer how our uncertainty relates to the absolute error, we report comparisons in Table \ref{tab:conferror}. To avoid making assumptions about the form of the error, we assess how well the uncertainty corresponds with error through Spearman's rank correlation $\rho$ (calculated on a large representative subset for computational reasons). This measures to what extent the two are related by a monotonic function. We additionally report the width of the intervals and how calibrated they are (measured by what fraction of the data in the test set falls within them). Most of our outputs show agreement between uncertain regions and large error regions. The ones with weakest correlation, field strength and $v_\textrm{los}$, already have an average interval width that is about the size of the bin: field is at bin-size and $v_\textrm{los}$ is about twice bin-size. We hypothesize that increasing the number of bins may improve the uncertainty modeling. The thermodynamic properties, on the whole, have substantially larger bin-sizes and worse calibration -- $\Delta \lambda_D$, $\eta_0$, and $S_0$ are all overly-narrow. Nonetheless, while this points to potential areas for improvement, the produced uncertainty has good correlation with error and is generally reasonably sized, with the exception of azimuth, which is extremely noisy for most of the Sun.

Finally, we report the rank correlation between our uncertainties and those produced by pipeline VFISV code (but which are used in downstream analysis). In addition to directly testing the quantities we do compare, good correspondence would indirectly validate the thermodynamic properties, for which there is no pipeline uncertainty. With the exception of field strength, Table \ref{tab:confconf} shows there is reasonably good correlation between the pipeline and emulated uncertainties; when one looks at higher-field regions, this correlation substantially improves. Pipeline VFISV uncertainties are derived solely from the local derivatives in parameter space, and represent lower-limits to the uncertainty of the value at any given data point.

\begin{deluxetable}{ccccc}[t]
\caption{Spearman's $\rho$ between Pipeline VFISV uncertainty and the width of the estimated 90\% confidence intervals. With the exception of field strength, there is reasonably good agreement between the system and the pipeline.}
\label{tab:confconf}
\tablehead{\colhead{Data Set}&\colhead{$B$} & \colhead{$\gamma$} & \colhead{ $\Psi$} & \colhead{$v_\textrm{los}$}}
\startdata
All &0.04 &0.30 &0.47 &0.26 \\
$>300$ \gauss &0.09 &0.73 &0.55 &0.49 \\
\enddata
\end{deluxetable}

\subsection{Comparison With Alternate Approaches} \label{comparison with alternate approaches}
\label{sec:experiments_ablations}

\begin{deluxetable*}{lccccccccc}[t]
\caption{Ablations of Different Losses and Models. We report results comparing MAE on the \textbf{2016-all} dataset.}
\label{tab:ablation}
\tablehead{ & \multicolumn{6}{c}{U-Net} & \multicolumn{2}{c}{MLP Model} & \colhead{Linear} \\
\colhead{Target} & \colhead{Proposed} & \colhead{1 Hot-W} & \colhead{1 Hot-UW} & \colhead{MSE} & \colhead{Prop.+BN} & \colhead{No-Meta} & \colhead{MSE} & \colhead{NLL} & \colhead{MSE}}
\startdata
Strength (\gauss) &9.7 &16.3 &20.2 &10.3 &26.6 &16.0 &26.3 &30.9 &32.9 \\
Inclination ($\,^\circ$) &0.58 &0.88 &0.92 &0.64 &1.45 &0.77 &1.51 &1.78 &2.29 \\
Azimuth ($\,^\circ$) &13.1 &13.7 &13.1 &11.4 &42.8 &20.9 &24.5 &21.5 &34.9 \\
\enddata
\end{deluxetable*}

We next quantitatively compare our approach with a number of alternate techniques, investigating which model decisions are important for performance, along with the relative trade-offs of various techniques. Six categories of models are considered, many of which are ablations (changes that remove a component individually to assess its standalone impact) of our proposed full model: 
\begin{enumerate}
    \item {\bf Alternate Classification Systems:} We ablate variations of our classification output, trying pairwise combinations of using the smoothed target compared to a one hot target. We found the one-hot technique to work poorly on rare values without applying a weight to the loss for rare classes. As such, results are presented as from {\it 1 Hot-W} (Weighted) as well as {\it 1 Hot-UW} (Unweighted).
    
    \item {\bf Regression:} We use the same network as proposed, but optimize a standard mean-squared error, and report this as {\it MSE}. Internally, rather than predict the raw values, the network predicts the z-scored values (i.e., zero-mean, unit variance), a process which is undone for evaluation. This compensates for the fact that native values from the VFISV output vary tremendously. 
    
    This tests whether our regression-by-classification, which comes with the ancillary benefits of enabling the identification of confident predictions, is detrimental to performance. Our goal in this experiment is to put the performance of the system in context, {\it not} to claim  that regression networks cannot outperform the classification network -- there may be alternate settings, losses, and schemes under which they may. 

    \item {\bf With Batch-Norm:} We use the same network as proposed, but incorporate Batch-Norm \diff{ \textcolor{darkgreen}{\cite[]{ioffe2015batch}}}, a standard practice, and report it as {\it Prop.+BN}. This comparison tests whether Batch-Norm is harmful to network performance.
    
    \item {\bf Without Auxiliary Channels:} We train the same network as proposed, but remove \diff{three of the auxiliary channels (latitude / longitude / on-disk flag),} and report it as {\it No-Meta}. This comparison tests whether this information is informative for the network.
    
    \item {\bf Multi-Layer Perceptron (MLP):} The VFISV pipeline was originally meant to have an initialization via a shallow (3-layer) fully-connected neural network per-pixel consisting of 30 neurons per layer. We compare with a modernized version of this network: we replace its activations with a ReLU to accelerate training convergence and implement the per-pixel fully-connected network via equivalent $1\times1$ convolutions to accelerate data processing. This comparison gives context to using a much deeper network.
    We refer to an MLP network trained with an MSE objective function as {\it MLP+MSE} and an MLP network trained with a negative-log-likelihood objective function as {\it MLP+NLL}.
    
    \item {\bf Linear Model:} There has been substantial work in using linear functions to perform Stokes inversions \diff{e.g., via principle component analysis (PCA) in \cite{socas2001fast}}. To test the performance of a linear model in the present context, we optimize a $1\times1$ convolutional neural network directly to targets, which is equivalent to learning linear weights for each of the 24 input channels as part of mapping them to outputs. We train this network with a MSE loss.
\end{enumerate}

We report results in Table \ref{tab:ablation} for only the magnetic field parameters in the interest of space and report ablations in terms of differences in loss function, inputs, and architecture.

\vspace{2mm}
\noindent {\bf Losses:} The smoothed target does substantially better on strength and inclination compared to one-hot schemes and the mean-square error, and does slightly worse (by only about $6\%$) on azimuth compared to one-hot. The azimuth error for the MSE-trained network is, however, substantially lower. We note though that the MSE-trained network comes with no uncertainty quantification.

\vspace{2mm}
\noindent {\bf Inputs:} Removing auxiliary information about the location on the disk from the network reduces performance on all targets. While VFISV is indeed a per-pixel process, 
%\textcolor{darkgreen}{the Milne-Eddington model incorporates spatial information related to observing angle (as it deems the optical depth as always along the viewing angle)}, and 
the full HMI processing pipeline includes position-dependent calibration information (e.g., in the instrument transmission profile and noise-level estimates).  Hence a decrease in performance without auxiliary information is not surprising.

\begin{figure}[t]
    \centering
    \includegraphics[width=\linewidth]{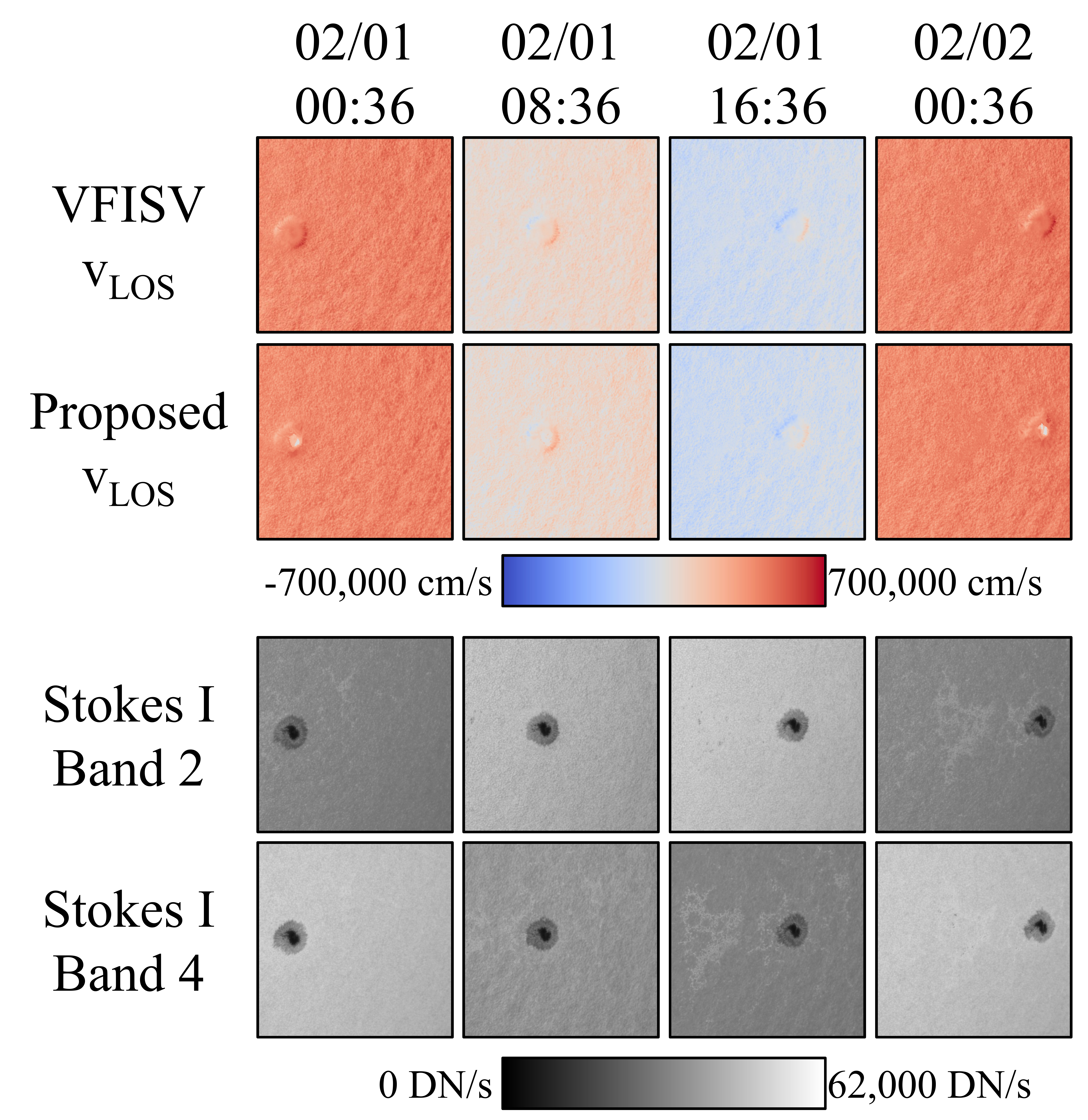}
    \caption{An illustration of where batchnorm can have harmful effects. The line of sight velocity (top row), here plotted at three evenly spaced times after 2016/01/01-00:36:00 TAI, depends on the velocity of the instrument. The amount of light falling into each bandpass (bottom two rows) varies due to this velocity. The varying average stokes band input, a clean signal for line-of-sight velocity, gets removed with input-dependent normalization or whitening.}
    \label{fig:batchnormfailure}
\end{figure}

\vspace{2mm}
\noindent {\bf Architectures:} While batch-norm is common practice in most networks that are trained on Internet images, adding it to the proposed network hurts, reducing performance to sometimes worse than a linear model. This is likely because many outputs depend on the {\it absolute} values of the input rather than their normalized/whitened versions: for example, the total amount of light in each passband is crucial to identifying the amount of Doppler shift. We illustrate this in Figure~\ref{fig:batchnormfailure}. Without this varying intensity signal, the network must rely on other, less effective, cues. With batchnorm, the input to the network would be improperly whitened and it would be difficult to model the variation seen in the top two rows. 

Substantially decreasing the capacity unsurprisingly has a negative impact on performance. As seen by Table \ref{tab:ablation}, our proposed model cuts the error rate by  $\sim 60\%$ for both field and inclination and by around  $\sim 45\%$ for azimuth compared to the style of network originally used in VFISV. Confirming the suggestions in \cite{Centeno2014}, a linear model does even worse: our model improves on MAE by $\sim 70\%$ for field and inclination and $\sim 60\%$ for azimuth over a linear MSE-trained model.

\subsection{Network Behavior Across a Time-Sequence} \label{network behavior over time}

% --- --- Cat Ears --- ---
\begin{figure}[t]
\centering
\includegraphics[width=\linewidth]{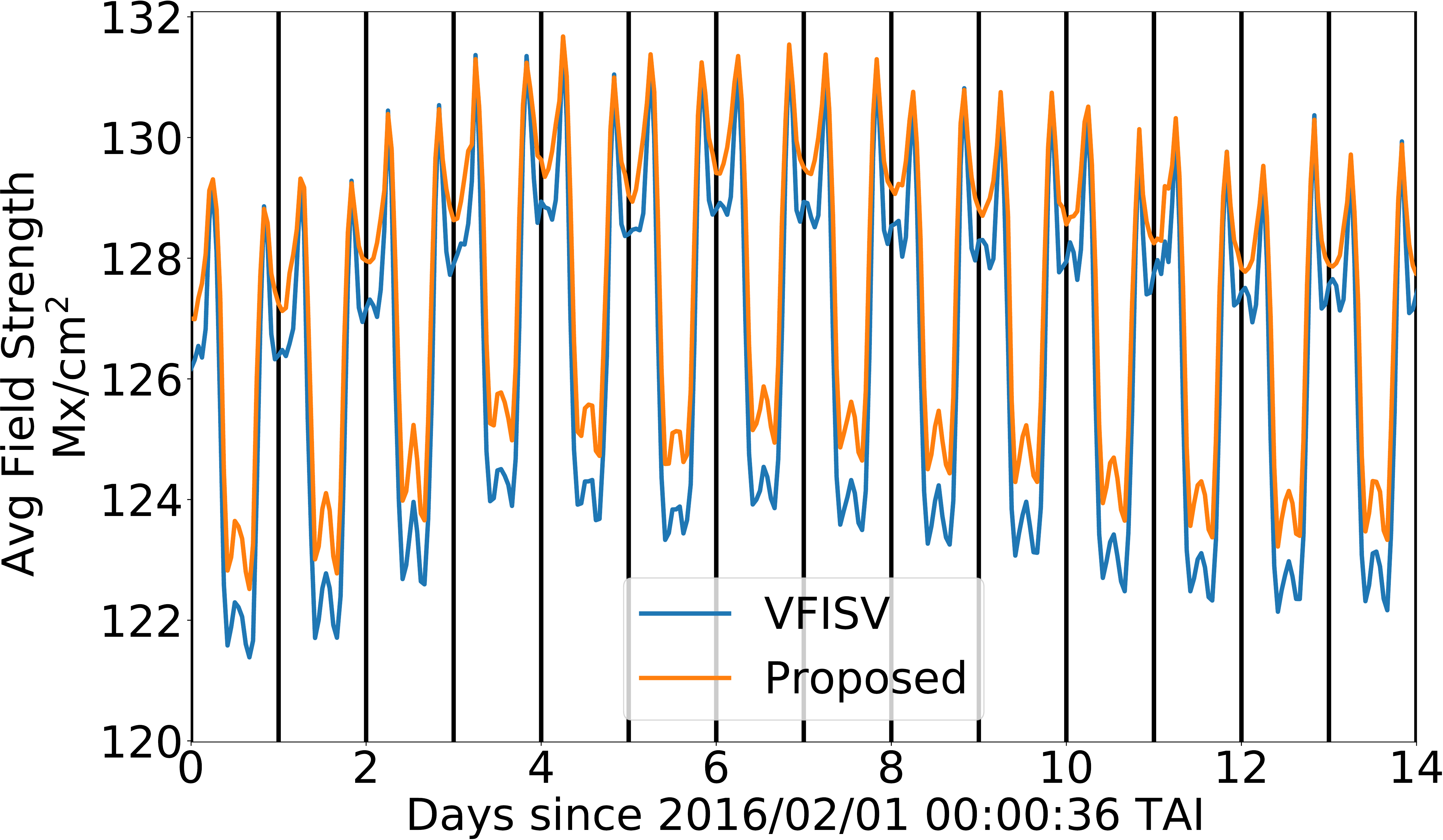} \\
\includegraphics[width=\linewidth]{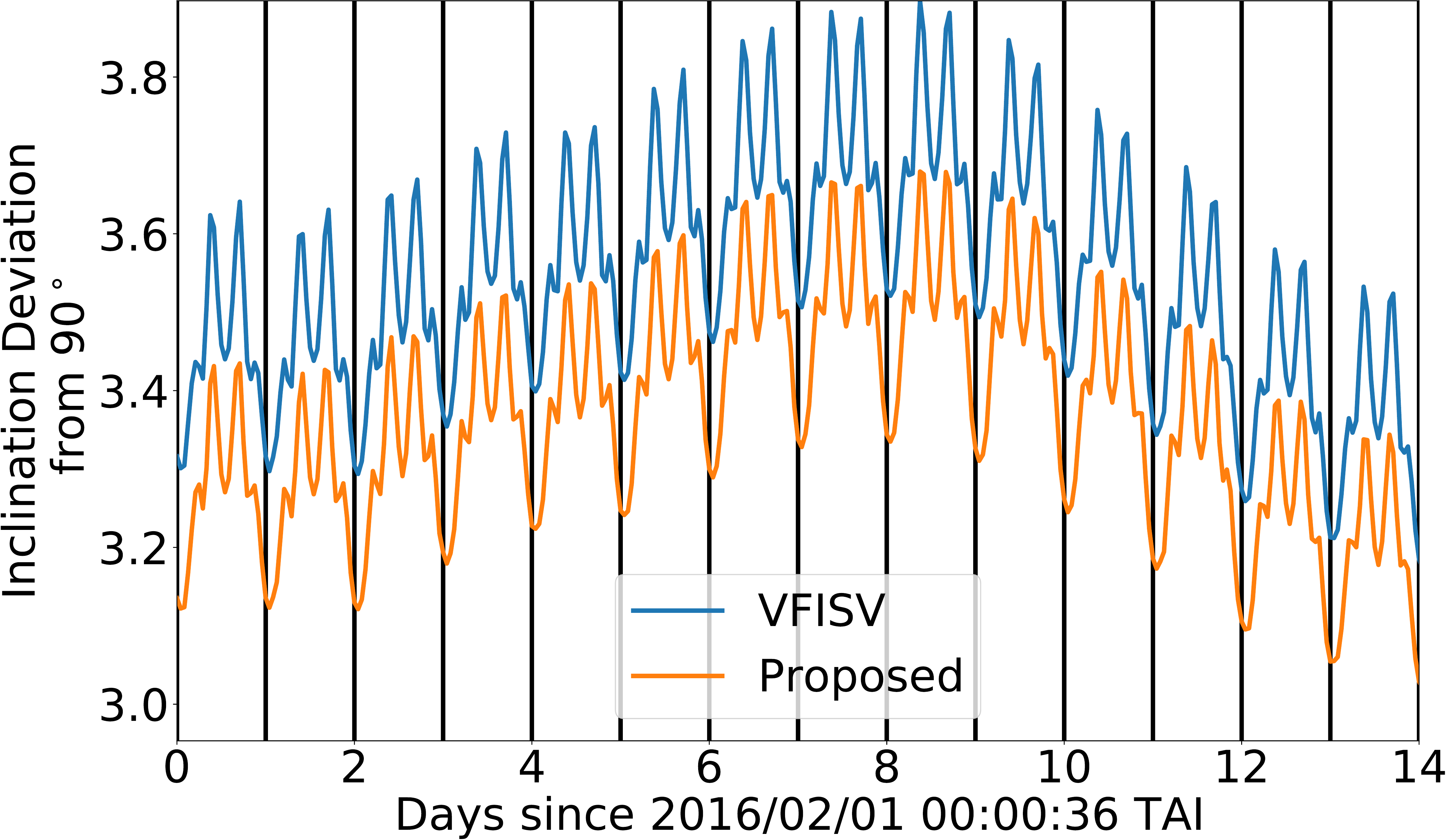} \\ ~ \\
\caption{Average on-disk field strength and deviation from horizontal as a function of time over a two week period with 24 hour periods separated by black vertical lines. The proposed system faithfully recreates known periodic behavior of the current \textit{SDO}/HMI pipeline.}
\label{fig:cat_ears}
\end{figure}
% --- --- Cat Ears --- ---

\begin{figure*}[t]
    \centering
    \includegraphics[width=\linewidth]{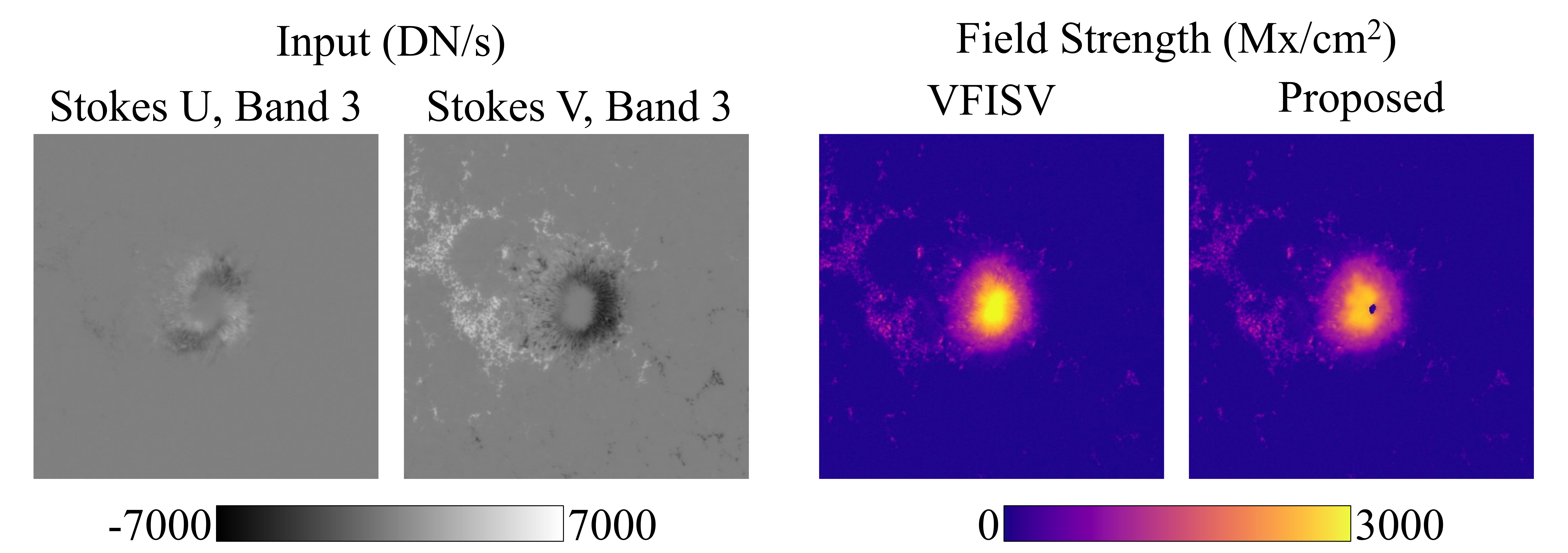}
    \caption{The system sometimes underpredicts in the spatial center (darkest part) of the largest sunspots. Here, the large Zeeman splitting and very low intensity make the near-line-center filtergrams appear similar to quiet sun for the CNN. Thus, a normally reliable polarization signal is removed from the information available to the training. Due to the rarity of this event, there are few pixels to provide a good alternate signal for the system to learn.}
    \label{fig:failure}
\end{figure*}

Finally, we evaluate how the network's outputs change in the temporal dimension. This is of interest since the network is trained solely on individual images, independent of a temporal requirement, and, as Eqn. \ref{eqn:objective} shows, is trained by minimizing an objective that considers each output pixel independently.

In particular, we analyze to what extent the network is able to capture known 24-hour oscillatory behavior of the pipeline by examining its output at a uniform, higher, hourly cadence over a month-long period.  We do this for two statistics: the average on-disk \diff{magnetic flux density} ($\frac{1}{n}\sum_{p} \textmd{B}_p$), and the average inclination distance from horizontal ($\frac{1}{n}\sum_{p} |\gammaB_p - 90^\circ|$). We plot these as a function of time over a two week period in Figure \ref{fig:cat_ears}. The results capture the periodicity of the oscillations, slightly offset by an average absolute error of $1.0$\,\gauss for field and $0.17^\circ$ for inclination. \diff{The flux density} error is 1.5\% of the width of a bin in our system, while inclination error is 7.5\% of the width of a bin.

We quantify how well our model predictions match the VFISV pipeline outputs with Spearman rank correlation $\rho$, which describes how well the two are related by a monotonic function (i.e., that could be applied post-hoc). Both are near \diff{unity}: 0.987 for field and 0.995 for inclination. Moreover, a per-hour-of-the-day additive correction estimated from a single day of data and applied to all remaining days drops the average absolute error to $0.17$\,\gauss and $0.01^\circ$. These experiments show that our system captures known periodic artifacts produced by the pipeline VFISV inversion as described in \cite{Hoeksema2014}. The alignment of around $\sim1$\,\gauss is surprising, given the field strength bin size of $\sim63$\,\gauss. This extreme correlation enables our approach to serve as an ultra-fast proxy for VFISV to aid the investigation into the root cause of the periodic oscillations.

\subsection{Failure Modes and Limitations} \label{sec:limitations}

We briefly describe some limitations of our current approach. Many of these limitations stem from extraordinarily rare events such as the \diff{strongest-polarization} pixels, especially those at the centers of large sunspots; performance may improve with more training data. Some instances of sub-optimal performance may stem from our use of identical models for all outputs, and may be fixed by tailoring the design decisions for particular targets.

\vspace{2mm}
\noindent {\bf Line-to-continuum ratio/$\eta_0$:} 
Some \textit{SDO}/HMI observations result in the Stokes inversion optimization objective having two distinct minima with a degeneracy involving $\eta_0$ and $B$. VFISV overcomes this with a term in its objective that prefers $\eta_0$ to be closer to a set value, $5$. Our method does not estimate $\eta_0$ well, which may be explained by this degeneracy.

\vspace{2mm}
\noindent {\bf Saturated Predictions:}
\diff{
Strong but sunspot-appropriate flux densities ($\sim$3000 \gauss) are rare in general and exceedingly so in the training set.  While the model shows good agreement for pixels in the $1000-2000$ \gauss range (which 0.07\% and 0.006\% of training pixels exceed, respectively), the model's predicted $B$ saturates at $\gtrsim$2750 \gauss (which fewer than $0.0003\%$ of training pixels exceed). We hypothesize the model may treat these rare-value points as similar to those in the more moderate-signal regime due to lack of training data. Future work could investigate avenues for improved performance in these strong-signal regimes including increased data samples, supplementing the training with synthesized data, or employing specialized sub-networks that address them exclusively. 
}

\vspace{2mm}
\noindent {\bf Large sunspots:}
Many active regions are precisely emulated by the model, as seen in Figures~\ref{fig:fulldisk},~\ref{fig:qual_main},~\ref{fig:random}. However, the very centers of some large sunspots can appear similar to quiet Sun regions when viewed solely in some of the input channels for the \textit{SDO}/HMI pipeline as seen in Figure~\ref{fig:failure}. Specifically, a lack of polarization signal (in raw counts) near line center can be explained by either strong Zeeman splitting or lack of magnetic field, notwithstanding the continuum intensity. Strongly Zeeman-split spectra are rare in the data, and so there are few samples for the training to consider when learning to look at the other wavelengths. The sunspot centers are thus sometimes predicted to have low field strength; this could likely be corrected by post-hoc processing or increased training data of sunspots.

\vspace{2mm}
\noindent {\bf Field Strength Uncertainties:}
While most of the model's uncertainties correlate well with both absolute errors and their counterparts in the VFISV pipeline \diff{(especially in the higher-flux regime), the uncertainties for the field strength parameter have lower correlation.} We hypothesize that it is difficult to express uncertainties smaller than the bin-size of the field outputs (for field, $\sim63$\gauss), and that narrower bins for field \diff{parameter} (and LOS velocity) may produce better uncertainties. In practice, the estimated field strength \diff{(flux density)} is itself a good fall-back predictor of the likely size of errors as shown by the slight growth in spread in the bivariate histograms.

\vspace{2mm}
\noindent {\bf Noisy and Circular Azimuth Angle:} 
The 90\% \diff{confidence} intervals for azimuth are, on average, much larger than other targets. Much of this is likely driven by the weak-polarization regions that make up the bulk of the Sun and hence dominate the training data, but have low SNR especially for the linear polarization.  We suspect an additional complication from the inherent $180^\circ$ ambiguity in the azimuth output from VFISV (and all other inversions) of Zeeman-based polarimetry.  In other words, the CNN cannot model the azimuth's circularity, the azimuth is treated like inclination, and the method is unaware that $0^\circ$ and $180^\circ$ are the same. Future work aiming for higher performance may wish to treat azimuth specially.

\vspace{2mm}
\noindent {\bf Multiple Networks:}
To avoid the issue of balancing all eight losses together in one equation, we train a single network per inversion output. An example of where cross-loss balancing may be an issue is the weighting of MSE components between the field strength parameter, ranging from 0 to 5000 \gauss, and the inclination angle parameter, ranging from 0 to 180 degrees. While using multiple networks maximizes accuracy and prevents failures due to poor loss scaling, it does increase runtime. Future work may wish to consolidate frequently-used subsets (e.g., field, azimuth, inclination) into a single network.

\vspace{2mm}
\noindent 
{\bf Varying Instrumental Calibration:}
\diff{
The reality of variations in instrumental characteristics and calibrationw
will lead to subtle differences between the training, validation, and test data.  In the present case, for example, the datasets chosen (Section\,\ref{sec:datasets}) were agnostic to an observing-sequence modification that HMI 
underwent mid-2016 \citep{Hoeksema_etal_2018}.  As such, were the training and testing datasets instead chosen specifically with this abrupt change under consideration, {\it i.e.} restricting all datasets to before or after, it is likely that error levels would have improved. A full investigation into the magnitude of such effects is beyond the scope of this paper, but this potential limitation does extend to all issues of unstable data quality -- be it through instrumental degradation, data-collection protocol, or varying calibration.
} \label{sec:experiments}
\section{Discussion and Conclusions}
We present a deep-learning approach for emulating the SDO/HMI pipeline VFISV Stokes inversion. \diff{The system emulates the pipeline well on an absolute basis (seen in Section \ref{standalone accuracy analysis}). The system usually produces estimates of uncertainty that are predictive of errors and agree well with uncertainties produced during the VFISV optimization procedure (seen in Section \ref{standalone confidence analysis}). We find that a careful design of the regression-via-classification problem, using relatively deep networks, and removing Batch-Norm \cite[]{ioffe2015batch} are all important for performance (seen in Section \ref{comparison with alternate approaches}). Finally, we show that the trained system faithfully recreates a periodic oscillation known to appear in \textit{SDO}/HMI pipeline outputs \cite[]{Hoeksema2014,schuck2016achieving}, as seen in Section \ref{network behavior over time}.}

The closest point of comparison between our results and other recent work is \cite{liu2020inferring} who applied a CNN to data from the Near Infrared Imaging SpectroPolarimeter at the Goode SolarTelescope. While we use similar base techniques of convolutional networks, there are several key differences in the data and methodology. First, our approach produces estimates of uncertainty compared to single point-estimates, the former are important for downstream applications, such as quickly identifying which estimates are likely to be correct and which are not. Second, our approach uses both Stokes components and auxiliary data in order to predict the inversion. We show that this auxiliary data improves results in our setting. Finally, our approach performs convolution over spatial resolution as opposed to a convolution over spectral dimensions. While \cite{liu2020inferring} report a slower speed to obtain field, inclination, and azimuth (one $720^2$ pixel in 50 seconds compared to three $4096^2$ in 15 seconds), we suspect hardware difference or implementation details drive this, since our network is also substantially deeper.

Direct quantitative comparison in terms of experimental results obtained is difficult due to the difference in the instruments and data. Nonetheless, even though GST/NIRIS has $10\times$ the spectral sampling, almost twice the spectral resolution, and $6\times$ the spatial resolution compared to SDO/HMI, our network produces comparable or better results. \cite{liu2020inferring} report MAEs of 134G/$6.5^\circ$/$13.2^\circ$ for field strength/inclination/azimuth in rectangular cutouts near two {\it unseen} active regions (with average predicted field strength $942$G and $62\%$ of pixels above $500$G). In unseen active regions (average field strength $1513$G, $99.98\%$ above $500$G), our system obtains MAEs of
108G/$2.5^\circ$/$10.7^\circ$. To better match the population from \cite{liu2020inferring}, we also computed tight bounding boxes around active region connected components bigger than $25^2$ pixels / $12.5^2$ arcseconds (average field strength $1185$G, $86.2\%$ pixels above $500$G). In these regions, our system obtains MAEs of $41$G/$1.1^\circ$/$5.3^\circ$.

Beyond quantitative accuracy measurements on overlapping targets, our classification experiments demonstrate a number of further contributions: we can additionally emulate kinematic and thermodynamic parameters, and our approach's uncertainty quantification usually carries meaningful information. Our ablation experiments extend these contributions, demonstrating both the detrimental impact of whitening features and the value of certain classification targets when applied to predicting physical quantities. Finally, our temporal experiments demonstrate that the system's {\it averaged} behavior both spatially across the disk and temporally, tracked {\it over weeks} behaves similarly to the \textit{SDO}/HMI pipeline output, which serves as further validation of our method. We see experiments like these, that go beyond pixel statistics, as critical to the future success and validation of deep-learning-based tools for solar physics.

From the analysis above, we conclude that our deep-learning approach provides two major enhancements to the standard pipeline for deriving photospheric magnetic fields from the HMI observations: speed and flexibility. Speed is generally a prerequisite for flexibility, but by itself, speed can dramatically enhance the effectiveness of the HMI data. Our approach has over two orders of magnitude faster time-to-solution than the present pipeline. \diff{Our speed-up originates from both the parallelism of GPUs and inference speed of CNNs, and using both together achieves the goal of real-time Stokes-vector inversion.} 

\diff{We see a number of important applications for an ultra-fast emulation of VFISV. Our method can serve as an initialization to the pipeline's optimization (replacing an earlier, now defunct, neural initialization). Functioning as a front-end to the pipeline, our method would provide an initial solution that is close to what the pipeline would derive, thereby speeding up the convergence of the pipeline's optimization and reducing resource usage.} Additionally, the increase in speed and faithful emulation of the oscillation artifacts may enable more rapid analysis of the source of these artifacts and lead to their correction. While our results are still azimuth-ambiguous and there is still room for improvement, we also see our work as a crucial step towards obtaining data faster, which may have many downstream impacts in space weather modeling. Looking towards the future, a far faster inversion process may aid in near-real-time forecasting and help in the direct driving of coronal MHD models, since recent work has suggested that the necessary cadence may be far faster than the 12-minute cadence of the HMI observations \citep{Leake17}. \diff{As a standalone system, our method can serve as a fast ``quick-look'' Stokes inversion for space weather forecasting applications when near real time data are needed before the definitive inversion is performed.}

In summary, we have presented in this paper a deep-learning approach for fast and accurate emulation of the HMI pipeline Stokes inversion module.
While our approach provides a more efficient way to produce existing information and does not produce new scientific models, it provides a first step towards advances like correcting hemispheric bias in HMI data, removing oscillation artifacts in HMI magnetograms, and extending solar magnetic field measurements with other observation modalities. In these cases, the prospect of correcting errors or making predictions without a corresponding detailed physical model has the potential to dramatically enhance a mission’s scientific value for solar and space research. Seen from this viewpoint, our ability to rapidly emulate the current pipeline is only a beginning.
 \label{sec:conclusions}
\acknowledgments
This work was supported by a NASA Heliophysics DRIVE Science Center (SOLSTICE) at the University of Michigan under grant NASA 80NSSC20K0600 and a Michigan Institute for Data Science Propelling Original Data Science grant.  GB and KDL also acknowledge NASA/GSFC grant 80NSSC19K0317.  All data used in this study are available from the Joint Science Operations Center (JSOC) at Stanford University, see \url{http://jsoc.stanford.edu/}. All relevant digital values used in the manuscript (both data and model) will be permanently archived at the U-M Library Deep Blue data repository, which is specifically designed for U-M researchers to share their research data and to ensure its long-term viability. Datasets will be assigned Digital Object Identifiers (DOIs) which will serve as identifiers for the data, enabling them to be cited in publications.
 \label{sec:acknowledgements}

\bibliography{main_fdlbsvicsh}{}
\bibliographystyle{aasjournal}

%% This command is needed to show the entire author+affiliation list when
%% the collaboration and author truncation commands are used.  It has to
%% go at the end of the manuscript.
%\allauthors

%% Include this line if you are using the \added, \replaced, \deleted
%% commands to see a summary list of all changes at the end of the article.
%\listofchanges

\end{document}